\documentclass[preprint]{aastex}
\usepackage{natbib}
\usepackage{lineno}
\usepackage{mathrsfs}
\usepackage{txfonts}

\usepackage{graphicx}
\usepackage{epsfig}

\begin{document}

\title{Observations of Milky Way Dwarf Spheroidal galaxies with the {\em Fermi}-LAT
  detector and constraints on Dark Matter models}

\author{
A.~A.~Abdo\altaffilmark{2,3}, 
M.~Ackermann\altaffilmark{4}, 
M.~Ajello\altaffilmark{4}, 
W.~B.~Atwood\altaffilmark{5}, 
L.~Baldini\altaffilmark{6}, 
J.~Ballet\altaffilmark{7}, 
G.~Barbiellini\altaffilmark{8,9}, 
D.~Bastieri\altaffilmark{10,11}, 
K.~Bechtol\altaffilmark{4}, 
R.~Bellazzini\altaffilmark{6}
% ...
}
% ...
\altaffiltext{2}{Space Science Division, Naval Research Laboratory, Washington, DC 20375, USA}
\altaffiltext{3}{National Research Council Research Associate, National Academy of Sciences, Washington, DC 20001, USA}
\altaffiltext{4}{W. W. Hansen Experimental Physics Laboratory, Kavli Institute for Particle Astrophysics and Cosmology, Department of Physics and SLAC National Accelerator Laboratory, Stanford University, Stanford, CA 94305, USA}
\altaffiltext{5}{Santa Cruz Institute for Particle Physics, Department of Physics and Department of Astronomy and Astrophysics, University of California at Santa Cruz, Santa Cruz, CA 95064, USA}
\altaffiltext{6}{Istituto Nazionale di Fisica Nucleare, Sezione di Pisa, I-56127 Pisa, Italy}
\altaffiltext{7}{Laboratoire AIM, CEA-IRFU/CNRS/Universit\'e Paris Diderot, Service d'Astrophysique, CEA Saclay, 91191 Gif sur Yvette, France}
\altaffiltext{8}{Istituto Nazionale di Fisica Nucleare, Sezione di Trieste, I-34127 Trieste, Italy}
\altaffiltext{9}{Dipartimento di Fisica, Universit\`a di Trieste, I-34127 Trieste, Italy}
\altaffiltext{10}{Istituto Nazionale di Fisica Nucleare, Sezione di Padova, I-35131 Padova, Italy}
\altaffiltext{11}{Dipartimento di Fisica ``G. Galilei", Universit\`a di Padova, I-35131 Padova, Italy}
% ...
%
%

\begin{abstract}
We report on the observations of 14 dwarf spheroidal galaxies with the \emph{Fermi} Gamma-Ray Space Telescope 
taken during the first 11 months of survey mode operations.  
The {\em Fermi} telescope, which is conducting an 
all-sky $\gamma$-ray survey in the 20 MeV to $>$300 GeV energy range, provides a new opportunity to test
particle dark matter models through the expected $\gamma$-ray emission produced by pair annihilation of 
weakly interacting massive particles (WIMPs).  Local Group dwarf spheroidal galaxies, the largest galactic 
substructures predicted by the cold dark matter scenario, are attractive targets for such indirect searches for 
dark matter because they are nearby and among the most extreme dark matter dominated environments.  
No significant $\gamma$-ray emission was detected above 100 MeV from the candidate dwarf galaxies. We determine upper limits to the $\gamma$-ray flux assuming both power-law spectra and representative spectra from WIMP annihilation. The resulting integral flux above 100 MeV is constrained to be at a level below around $10^{-9}$ photons cm$^{-2}$s$^{-1}$.
Using recent stellar kinematic data, the $\gamma$-ray flux limits are combined
  with improved determinations of the dark matter density profile in 8 of the 14 candidate dwarfs to place limits on the pair annihilation
  cross-section of WIMPs in several widely studied
  extensions of the standard model, including its supersymmetric extension and other models that received recent attention.  
With the present data, we are able to rule out
large parts of the parameter space where the thermal relic density is
below the observed cosmological dark matter density and WIMPs (neutralinos here) are
dominantly produced non-thermally, e.g. in models where supersymmetry
breaking occurs via anomaly mediation.
The $\gamma$-ray limits presented here also  
constrain some WIMP models proposed to explain the \emph{Fermi} and PAMELA
  $e^+e^-$ data, including low-mass wino-like neutralinos and models with TeV masses pair-annihilating into muon-antimuon pairs.
\end{abstract}

\section{Introduction}

A wealth of experimental evidence and theoretical arguments have 
accumulated in recent years in favor of the existence of some form of 
non-baryonic cold dark matter (CDM) to explain the observed large-scale structure in the Universe.
According to the most recent estimates,
CDM constitutes approximately one-fourth of the total energy density of the
Universe.  However, very little is known about the underlying nature of this
dark matter, despite
the efforts of high-energy physicists, astrophysicists and cosmologists over many years, and it 
remains one of the most fascinating and intriguing issues in present day physics.
One appealing possibility is that CDM consists of 
a new type of weakly interacting massive particle (WIMPs), 
that are predicted to exist in several theories beyond the Standard Model of 
particle physics. Such WIMPs typically have pair-annihilation cross sections that, for their natural mass range (between a few GeV and a few TeV), drive a thermal relic abundance in the same ballpark as the inferred amount of cosmological dark matter. 
Pair-annihilation (or decay) would also occur today, yielding, 
among other particle debris like energetic neutrinos, (anti-)protons and electron-positron pairs, a significant flux of high-energy $\gamma$-rays. If the dark matter is meta-stable, its decay products would also produce potentially detectable $\gamma$ rays. While the results we present here would constrain this type of scenario, we assume here that the dark matter particle is stable.

Cosmological N-body simulations of structure formation show that the
dark matter halos formed by WIMPs are not smooth and have a large
number of bound substructures (subhalos) whose numbers increase with
decreasing mass \citep{2005Natur.435..629S,2008JPhCS.125a2008K,2005Natur.433..389D}.
Dwarf spheroidal galaxies (dSphs), the largest galactic substructures predicted by the CDM
scenario, are ideal laboratories for indirect searches for dark matter, through the observation of 
dark matter annihilation (or decay) products, for the following reasons.
The mass-to-light ratios in dSphs can be of order
$100-1000$ (see Table \ref{dwarfList}), showing that they
are largely dark matter dominated systems. Therefore, the stars serve as tracer particles
in the dark matter gravitational potential and the dark matter
distribution in these dwarfs may be constrained using stellar kinematics. 
In addition, dSphs are expected to be relatively free
from $\gamma$-ray emission from other astrophysical sources as they
have no detected neutral or ionized gas, and little or no recent star
formation activity
\citep{1998ARA&A..36..435M,2003ApJ...588..326G,2009ApJ...696..385G}, thus simplifying the interpretation of a gamma-ray excess that would be detected in the direction of a dSph. 
In addition, the Sloan Digital Sky Survey \citep[SDSS, see][]{2000AJ....120.1579Y} has led, in recent years,
 to the discovery of a new population of ultrafaint Milky Way satellites, comprising about
as many (new) objects as were previously known
\citep{2005AJ....129.2692W,2006ApJ...650L..41Z,2007ApJ...656L..13I,2007ApJ...662L..83W,
2007ApJ...654..897B}. This new population of extremely low-luminosity, but dark matter dominated galaxies could in particular 
be very interesting for indirect dark matter searches \citep{2008ApJ...678..614S}, 
especially with the \emph{Fermi} Gamma-Ray Space Telescope.

\emph{Fermi} is a new generation space observatory, which was successfully launched on June 11th, 2008, and has been operating
 in nominal configuration for scientific data taking since early 
August 2008\footnote{For more details, see the Fermi website at http://fermi.gsfc.nasa.gov/}. 
Its main instrument, the Large Area Telescope (LAT), is designed to explore the high-energy $\gamma$-ray sky in the $20$~MeV to $>300$~GeV energy range, with unprecedented angular resolution and sensitivity.
Several studies have been performed to determine 
the sensitivity of the LAT to dark matter annihilation signals \citep[e.g.][]{2008JCAP...07..013B} and
the \emph{Fermi}-LAT collaboration is currently exploring several potentially complementary
search strategy for $\gamma$-ray emission from dark matter.
We focus here on the search for a $\gamma$-ray signal in the direction of
a selection of 14 dSphs. The criteria for this selection, together with
the description of the \emph{Fermi}-LAT data analysis, are presented in \S
\ref{data}. We determine upper limits to the $\gamma$-ray flux employing both power-law spectra with spectral indexes in the range between 1 and 2.4 (\S \ref{powerlaw}), and spectra resulting from the annihilation of several representative WIMP models, for various WIMP masses (\S \ref{DMconstraints}). 
To turn these results into limits on the WIMP pair annihilation cross section, we restrict our focus
to a subset of 8 dSphs that are associated with stellar data of good
enough quality to allow for an accurate modeling of their dark matter
content. We then present an updated determination of the assumed Navarro-Frenk-White dark matter density profile,
using a Bayesian analysis and up-to-date stellar velocity data (\S \ref{profiles}). We show in \S \ref{DMconstraints} results for WIMP models in the context of minimal supergravity, of a general weak-scale parameterization of the minimal supersymmetric standard model, of the minimal anomaly-mediated supersymmetry breaking scenario, and of universal extra-dimensions. Finally, in \S \ref{mumu}, we discuss these constraints in the context of dark matter annihilation models that fit the PAMELA and \emph{Fermi} $e^+e^-$ data, putting special emphasis on the effect that a possible contribution of secondary $\gamma$-ray emission from IC scattering has on them.
Our main conclusions are summarized in \S \ref{conclusion}.

\begin{deluxetable}{cccccccccc}
\tabletypesize{\scriptsize}
\tablecaption{Properties of the dwarf spheroidals used in this study.}
\tablewidth{0pt}
\tablehead{ 
\colhead{Name} & \colhead{Distance} & \colhead{year of}  & \colhead{M$_{1/2}$/L$_{1/2}$} & \colhead{l} & \colhead{b} & \colhead{Ref.}\\ 
               & \colhead{(kpc)}    & \colhead{discovery}& \colhead{ref. 8}	         &	       &             & 
}
\startdata
Ursa Major II      & 30$\pm$ 5  & 2006	    & $4000^{+3700}_{-2100}$        & 152.46 &  37.44 & 1,2\\
Segue 2            & 35         & 2009      & $650$ & 149.4  & -38.01 & 3\\
Willman 1          & 38$\pm$ 7  & 2004	    & $770^{+930}_{-440}$            & 158.57 &  56.78& 1\\
Coma Berenices     & 44$\pm$ 4  & 2006	    & $1100^{+800}_{-500}$          & 241.9  &  83.6 & 1,2\\
Bootes II          &  46        & 2007      & $18000$??                   & 353.69 & 68.87  & 6,7\\
Bootes I           &  62$\pm$3  & 2006	    & $1700^{+1400}_{-700}$                 & 358.08 & 69.62  & 6\\
Ursa Minor         & 66$\pm$ 3  & 1954	    & $290^{+140}_{-90}$           & 104.95 &  44.80& 4,5\\
Sculptor           & 79$\pm$ 4  & 1937	    & $18^{+6}_{-5}$           & 287.15 & -83.16 & 4,5\\
Draco              & 76$\pm$ 5  & 1954	    & $200^{+80}_{-60}$           &  86.37 &  34.72 & 4,5,9\\
Sextans            & 86$\pm$ 4  & 1990	    & $120^{+40}_{-35}$            & 243.4  &  42.2 & 4,5\\
Ursa Major I       & 97$\pm$4   & 2005	    & $1800^{+1300}_{-700}$         & 159.43 & 54.41  & 6\\
Hercules           & 132$\pm$12 & 2006	    & $1400^{+1200}_{-700}$          &  28.73 &  36.87 & 6\\
Fornax             & 138$\pm$ 8 & 1938	    & $8.7^{+2.8}_{-2.3}$         & 237.1  & -65.7  & 4,5\\
Leo IV             & 160$\pm$15 & 2006	    & $260^{+1000}_{-200}$          & 265.44 & 56.51  & 6\\
\enddata
\tablecomments{M$_{1/2}$/L$_{1/2}$ is the ratio of the total mass within the 3D half-light
radius to the stellar luminosity within the same radius from \citet{Wolf09}. 
The problematic result for Bootes II is further discussed in the text. Uncertainties in the determination of this
mass-to-light ratio (unavailable for Bootes II and Segue 2) arise from the errors in both M$_{1/2}$ and L$_{1/2}$, but they do not change the qualitative conclusion that these dSphs are dark matter dominated even within their stellar extent. 
References: 
(1) 
\cite{2008ApJ...678..614S},
(2) 
\cite{2007ApJ...670..313S},
(3) 
\cite{2009MNRAS.397.1748B},
(4) 
\cite{2008ApJ...672..904P},
(5) 
\cite{1998ARA&A..36..435M},
(6) 
\cite{2008ApJ...684.1075M},
(7) 
\cite{2009ApJ...690..453K}
(8) 
\cite{Wolf09}
(9) 
\cite{2004AJ....127..861B}
}
\label{dwarfList}
\end{deluxetable}
%

\section{\emph{Fermi}-LAT observations and data analysis  }
\label{data}

The LAT is an electron-positron pair conversion telescope 
sensitive to photon energies from $20$~MeV to $>300$~GeV \citep{LATpaper} .
It is made of 16 towers each comprising a tracker and a calorimeter underneath.
The tracker is made of silicon-strip active planes interleaved with tungsten foils, and 
is responsible for the conversion of the incident photon into an electron-positron pair and for the tracking of the 
latter charged particles. The energy of the photon is mainly estimated from the light deposited in the CsI(Tl) scintillators
that constitute the calorimeter. Finally, an anticoincidence detector, made of more than 100 plastic scintillators, covers the 16 towers
in order to be able to reject the charged particle background.
The LAT nominally operates in a scanning mode observing the whole sky
every 3 hours, the resulting overall coverage of the sky being fairly uniform.
The analysis described here uses data taken in this mode 
during the first eleven months of sky survey operation (August
4 2008 to July 4 2009). 

The dSphs that have been considered for this work 
are listed in Table \ref{dwarfList}. They were selected based on their proximity, 
high Galactic latitude and their dark matter content, which have been estimated from the most recent resolved stellar velocity measurements. 
In particular, Carina and Sagittarius were discarded based on the fact that they are at low latitude ($|b|<30^\circ$), and thus subject to potentially large systematics due to uncertainties associated with the modeling of the Galactic Diffuse emission as seen by the LAT. 
In addition, Segue~1 is a controversial case : while \citet{2009ApJ...692.1464G} concluded that this dwarf is the most promising satellite for indirect dark matter detection, this claim was challenged by \citet{2009MNRAS.398.1771N}, who contend that Segue~1 is a star cluster stripped early on from the Sagittarius galaxy. As new stellar data on Segue~1 are currently being analyzed~\citep{gehaTeVPA}, which may greatly improve the still uncertain measurements of the density profile, we defer its study to an upcoming dedicated publication.
Finally, Bootes II is modeled based on the published data on 5 stars. The result that we obtain for the mass to luminosity ratio is unrealistic, and could mean that something is wrong with the stellar membership of this system or that it is simply unbound. Nevertheless, more data exist and are currently being reduced, and proposals are under way to increase the stellar dataset for this object. As a result, we keep Bootes II in the present analyis, since in the future it may prove to be one of the best candidate dwarfs. Finally, in section \ref{models}, we use a subset of 8 dwarfs that have robust stellar kinematic data to further constrain models.

The data reduction makes use of the standard LAT ground processing 
and background rejection scheme described in \cite{LATpaper}, and we consider only 'Diffuse' class events, which
have the highest probability of being photons. Throughout, we use the {\tt Fermi ScienceTools version v9r15}, 
a software package dedicated to the \emph{Fermi}-LAT data analysis\footnote{http://fermi.gsfc.nasa.gov/ssc/data/analysis/software/}.
First, observations toward each dSph are extracted in 14 regions of interest (ROI), by 
keeping events that have a reconstructed direction of incidence at most 10$^\circ$ away from the dwarf position. This ROI accomodates the large point spread function (PSF) of the LAT at low energy. Indeed, the LAT PSF, which depends on the photon energy and angle of incidence, can be approximated by the function $0.8\,(E/1$GeV$)^{-0.8}$\,deg, yielding $\sim5^o$ at 100 MeV. 
Next, in order to avoid calibration uncertainties at low energy and background contamination at high energy, 
we apply a cut on the reconstructed energy $E$: $E>100$~MeV and $E<50$~GeV.  Here we employ a somewhat conservative cut at high energies to reduce the background, but work is ongoing within the \emph{Fermi} collaboration to develop an improved event selection which will have less high energy background contamination (Abdo et al., in preparation).
To avoid albedo $\gamma$-ray contamination, we also select Good Time Intervals (GTIs)
when the entire ROI is above the albedo horizon of the Earth (105$^\circ$ below the zenith).
Furthermore, the Earth limb appears at a zenith angle of 113$^\circ$ from Fermi's orbit. Thus,
 time periods during which the spacecraft rocking angle 
(the angle between LAT normal and Earth-spacecraft vector)
is larger than 43$^o$ are excluded as an additional
guard against Earth albedo $\gamma$-ray contamination. 

The resulting dataset is analyzed with a binned 
likelihood technique~\citep{1979ApJ...228..939C,EGRET_like}, implemented
in the \texttt{ScienceTools} as the \texttt{gtlike} task. 
{\tt gtlike} uses the maximum likelihood statistic to fit the data to a spatial \emph{and} spectral source model.
Because the numbers of photons from sources near the detection limits are fairly small, 
{\tt gtlike} calculates a likelihood function based on the Poisson
probability using the source model folded through the 
LAT  Instrument Response Functions (IRFs)  
\footnote{http://www-glast.slac.stanford.edu/software/IS/glast$\_$lat$\_$performance.htm}
to provide the expected model counts. This analysis relies on version P6$\_$V3 of the IRFs.
For each ROI, the source model includes all the preliminary 11 month LAT catalogue sources
 within a 10$^o$ radius of each dwarf. Following the analysis procedure used in the development of the LAT catalogue, 
these sources are modeled as point-like with power-law spectra, a reasonable approximation in the absence of dedicated studies for most of them. Furthermore, the positions and spectral parameters of these sources are being kept fixed during the fitting procedure.
It also includes the models currently advocated by the LAT collaboration\footnote{Detailed description can be found under `Model Description' at the following web page : http://fermi.gsfc.nasa.gov/ssc/data/access/lat/BackgroundModels.html}
 for the Galactic diffuse emission 
and for the corresponding isotropic component (which accounts for any extragalactic diffuse 
emission and any residual charged background contamination). 
Their independent normalizations are kept free during the fit in order
 to account for uncertainties in modeling these diffuse components.
Finally, the dSphs are modeled as point sources, their localization being kept fixed during the fit. 
Given the limited angular resolution of the LAT and the limited statistics, 
the point source approximation is reasonable for all the selected dwarfs. 
To model the source spectra, we employ two strategies: we employ model-independent power-law spectra, with a wide range of spectral indexes from 1 to 2.4, discussed in \S \ref{powerlaw}, as well as a collection of motivated and representative $\gamma$-ray spectra from WIMP pair annihilation, for wide ranges of masses and for several WIMP models (see \S \ref{DMconstraints}).

%
\subsection{Power-law modeling}\label{powerlaw}
For a power-law spectrum, the differential flux is written as
\begin{equation}\label{pwl}
 \frac{dN}{dEdAdt} = N_0\left(\frac{E}{E_0}\right)^{-\Gamma}\ ,
\end{equation}
where $E$ is the photon reconstructed energy in the restricted energy range 100~MeV to 50~GeV, and $E_0$ is an arbitrary energy scale set to 100 MeV. Such a model has two unknown parameters, the photon index $\Gamma$ and the normalization parameter $N_0$. In this analysis, we fix $\Gamma$ to five possible values,
 $\Gamma=\,$1, 1.8, 2.0, 2.2 and 2.4. While the last four indices
provide constraints on standard astrophysical source spectra, the very hard index of $\Gamma=\,1$ is motivated by the dark matter annihilation models in \cite{2009PhRvD..80b3506E}. $N_0$ is fitted to the data in each ROI separately, together with the isotropic and Galactic diffuse normalizations. The best fit values and corresponding errors of these three parameters, for the case $\Gamma=2$,  are gathered in Table~\ref{fitResult_PWL}.
In the third column, the errors on $N_0$ are several orders of magnitude larger than the fitted values, which means that the latter are compatible with zero. We thus conclude that no significant signals from the direction of the selected dSphs are detected. The same conclusion is reached for the other values of the photon index $\Gamma$.
\begin{deluxetable}{lccc}
\tabletypesize{\scriptsize}
\tablecaption{Results for the fit of the three normalization parameters in each ROI.}
\tablewidth{0pt}
\tablehead{
           \colhead{dSph} & \colhead{Galactic foreground}   & \colhead{Isotropic component}    & \colhead{$N_0$ ($\times10^{-5}$)}
}
\startdata
  Ursa Major II & 0.89 $\pm$ 0.07 & 0.97 $\pm$ 0.06 & $5.8 \times 10^{-09}$ $\pm$ $9.29 \times 10^{-06}$\\
        Segue 2 & 1.01 $\pm$ 0.04 & 1.03 $\pm$ 0.06 & $2.3 \times 10^{-15}$ $\pm$ $5.22 \times 10^{-10}$\\
      Willman 1 & 0.44 $\pm$ 0.25 & 1.07 $\pm$ 0.07 & $7.3 \times 10^{-05}$ $\pm$ $1.21 \times 10^{-04}$\\
  Coma Berenice & 0.90 $\pm$ 0.15 & 1.06 $\pm$ 0.06 & $2.0 \times 10^{-13}$ $\pm$ $4.54 \times 10^{-09}$\\
      Bootes II & 0.96 $\pm$ 0.14 & 1.19 $\pm$ 0.08 & $3.4 \times 10^{-12}$ $\pm$ $2.69 \times 10^{-08}$\\
       Bootes I & 0.80 $\pm$ 0.13 & 1.26 $\pm$ 0.08 & $1.2 \times 10^{-12}$ $\pm$ $2.32 \times 10^{-08}$\\
     Ursa Minor & 0.50 $\pm$ 0.11 & 1.13 $\pm$ 0.06 & $4.2 \times 10^{-13}$ $\pm$ $6.15 \times 10^{-09}$\\
       Sculptor & 0.53 $\pm$ 0.15 & 1.03 $\pm$ 0.06 & $1.2 \times 10^{-04}$ $\pm$ $1.40 \times 10^{-04}$\\
          Draco & 0.70 $\pm$ 0.09 & 1.08 $\pm$ 0.06 & $2.1 \times 10^{-11}$ $\pm$ $5.68 \times 10^{-08}$\\
        Sextans & 1.00 $\pm$ 0.10 & 1.09 $\pm$ 0.06 & $6.6 \times 10^{-12}$ $\pm$ $3.15 \times 10^{-08}$\\
   Ursa Major I & 0.71 $\pm$ 0.27 & 1.02 $\pm$ 0.07 & $2.9 \times 10^{-14}$ $\pm$ $1.84 \times 10^{-09}$\\
       Hercules & 0.93 $\pm$ 0.07 & 1.27 $\pm$ 0.09 & $1.1 \times 10^{-14}$ $\pm$ $2.56 \times 10^{-09}$\\
         Fornax & 1.01 $\pm$ 0.16 & 0.86 $\pm$ 0.06 & $1.1 \times 10^{-15}$ $\pm$ $2.31 \times 10^{-10}$\\
         Leo IV & 0.94 $\pm$ 0.11 & 1.23 $\pm$ 0.06 & $2.1 \times 10^{-11}$ $\pm$ $8.46 \times 10^{-08}$\\ 
\enddata
\tablecomments{Results are shown for the case $\Gamma=2$. Errors are statistical only. $N_0$ is the power-law prefactor in eq.(\ref{pwl}).}
\label{fitResult_PWL}
\end{deluxetable} 

\begin{figure}
\begin{center}
\includegraphics*[width=8cm,height=8cm,angle=270]{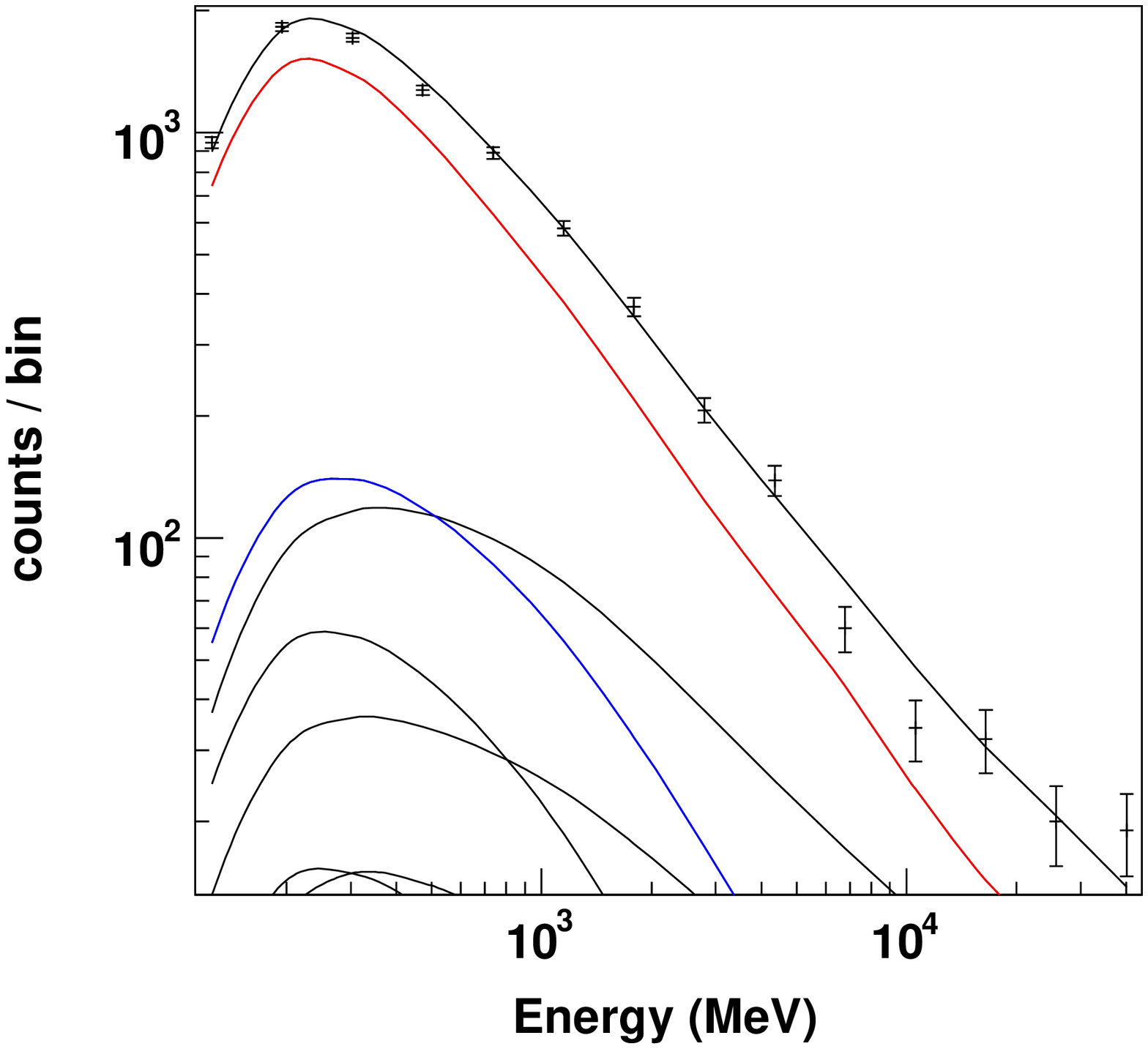}
\includegraphics*[width=8cm,height=8cm,angle=270]{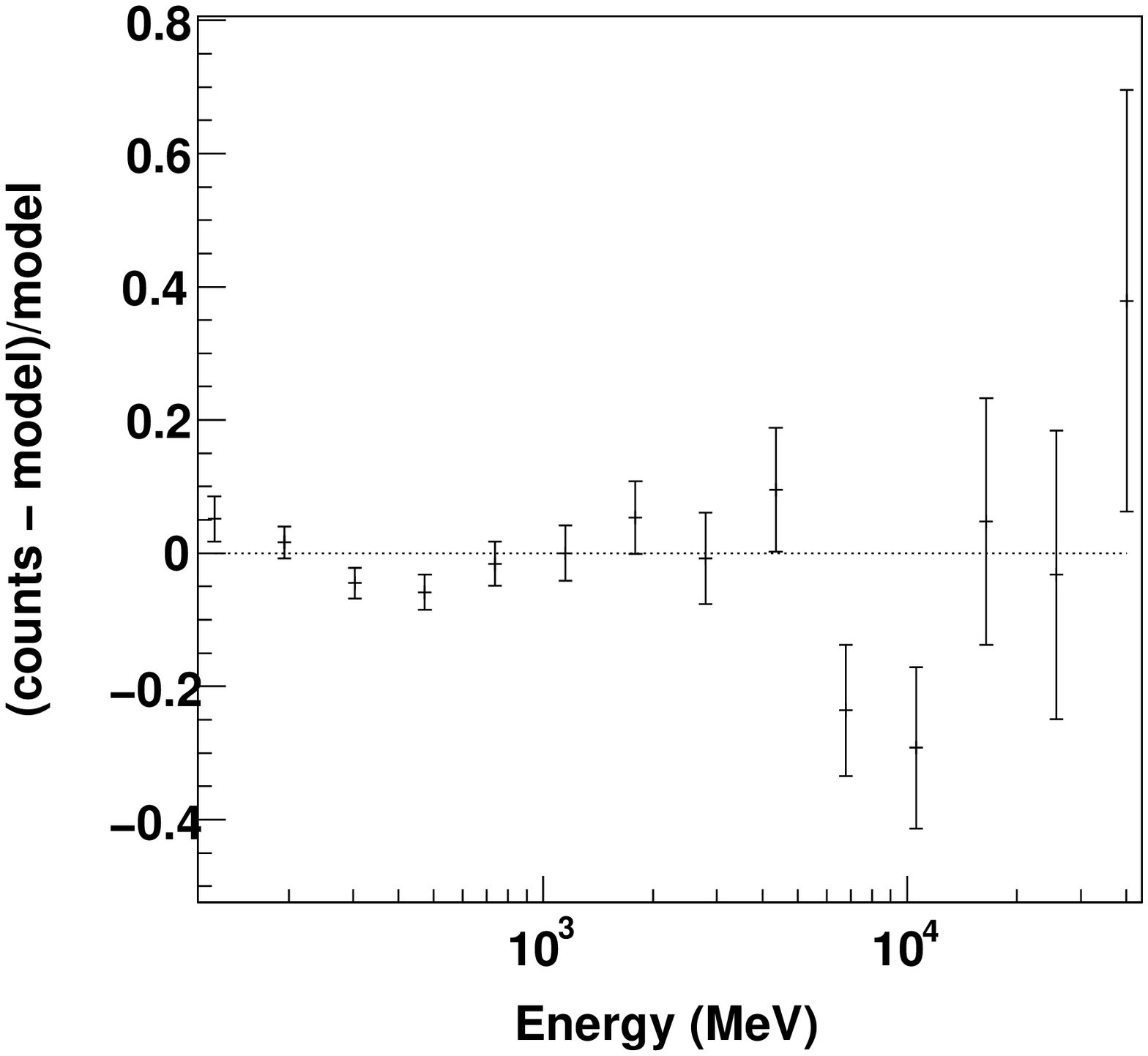}
\includegraphics*[width=8cm,height=8cm,angle=270]{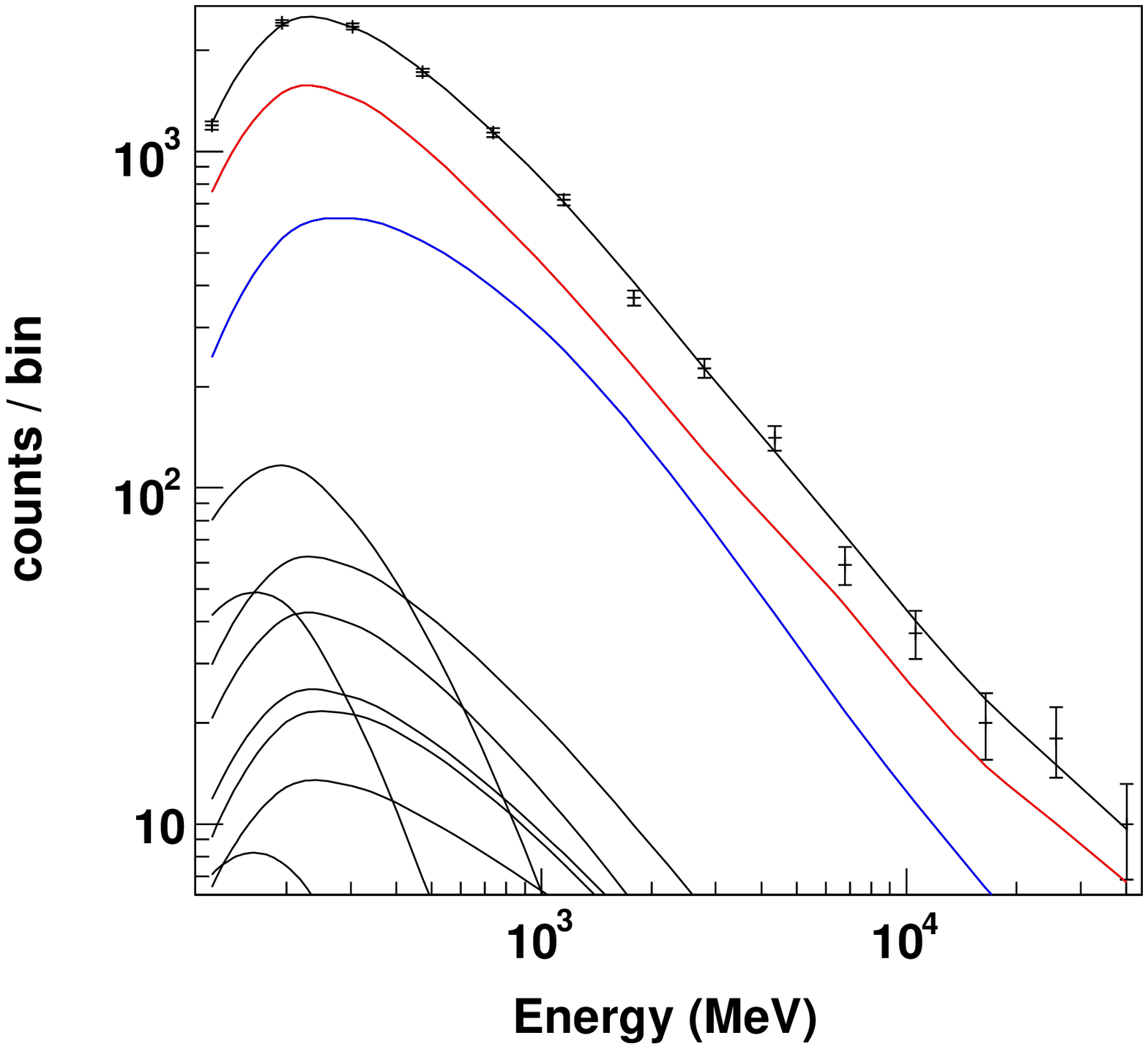}
\includegraphics*[width=8cm,height=8cm,angle=270]{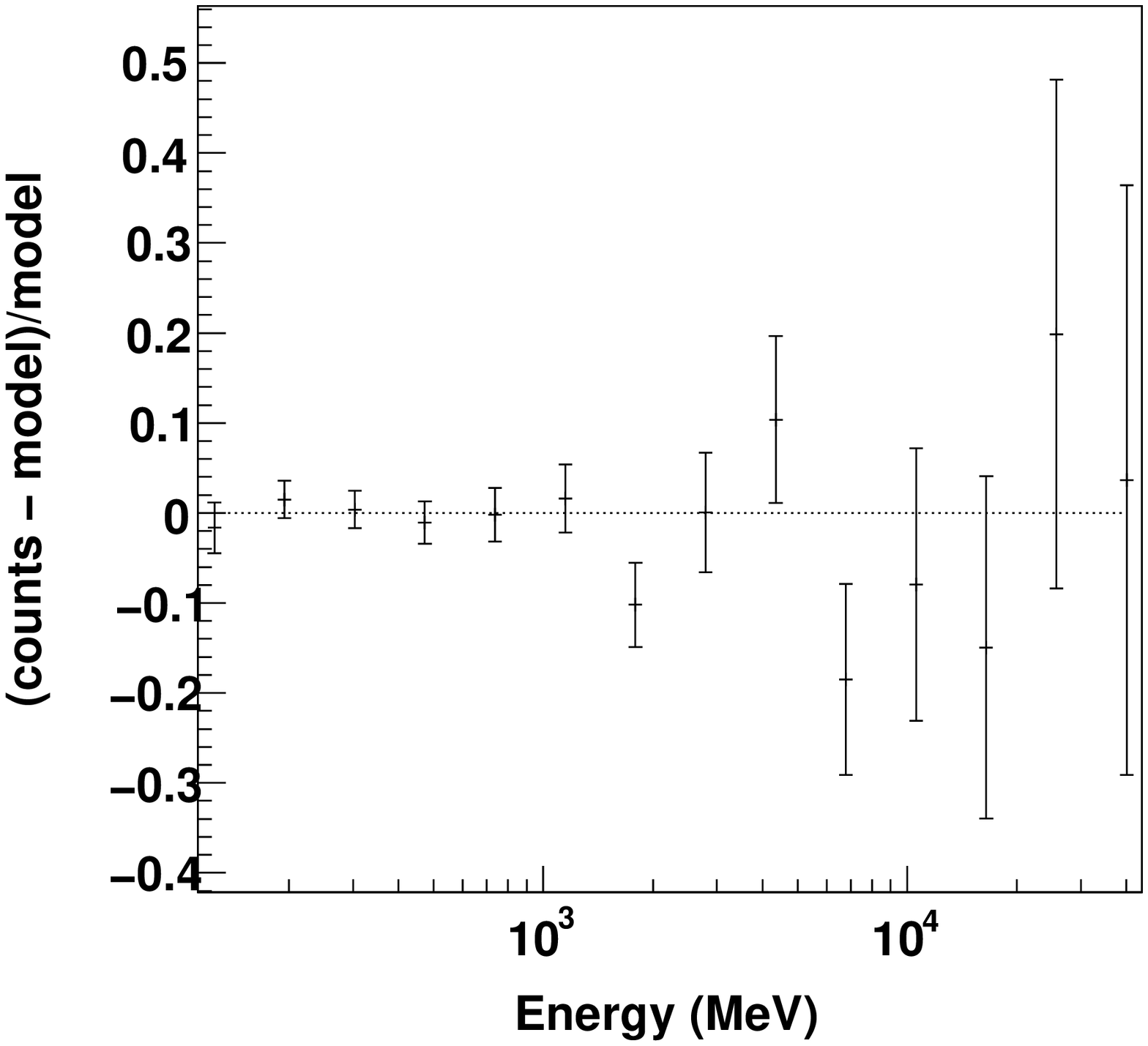}
\end{center}
\caption{\label{fitresid}
Spectral fits to the counts (left panels) and the corresponding residuals (right panels) for the ROIs
around two dwarf spheroidal galaxies, Willman 1 (top panels) and Draco
(bottom panels). The lines in the spectral plots (left panels) are
point sources (black), the Galactic diffuse component (blue) and the
isotropic component (red). The black line
overlaid to the data points is the best-fit total spectrum in the
respective ROIs. The best-fit power-law models (with $\Gamma=2$ here) for the dwarfs are below the lower bound
of the ordinates. Willman 1 is the worst residual obtained in our sample, while Draco is illustrative of the fit quality for most ROIs.
}
\end{figure}

We expect variations across the ROIs around the ideal value of 1 for the normalizations of the two diffuse components, due to possibly slightly different exposure, statistical fluctuations, and inaccuracy of the underlying spatial model and spectral shape. Nevertheless, the fit results for the isotropic component remain close to 1, while the deviations for the Galactic diffuse component are somewhat larger, which is expected as spatial inaccuracies of the model are expected to vary from one ROI to another. Fig.~\ref{fitresid} shows examples of the spectral fits to the data and the
residuals of these fits for Willman 1, which has the largest fit
residuals, and Draco, which has residuals typical of most of the fits.

Flux upper limits are then derived, based on a profile likelihood technique~\citep{Bartlett53, Rolke05}. In this method, the normalization of the power law source representing the dwarf galaxy is scanned away from the fitted minimum while the remaining two free parameters (the normalizations of the two diffuse backgrounds) are fit at each step.  The scanning proceeds until the 
difference of the logarithm of the likelihood function reaches 1.35, which corresponds to a one-sided 95\% confidence level (C.L.). Extensive work to test this method, using Monte Carlo simulations as well as bootstrapping the real
data, indicate that profile likelihood method as implemented in the Science Tools is slightly overcovering the expected confidence level\footnote{The upper limits derived from the tests covered
98\% of the trials instead of the 95\% required.}. As a result, we believe the ULs presented in this paper are
conservative.

In Table~\ref{dwarf_DM_UL}, we report flux upper limits in two different energy ranges 
(from 100MeV to 50GeV and from 1GeV to 50GeV), for the five different values of $\Gamma$.
 As expected, the \emph{Fermi}-LAT limits are stronger for a hard spectrum which predicts relatively more photons at the dwarf location at high energy.  At higher energies, the LAT PSF is significantly reduced while the effective area is significantly higher; in addition, the diffuse backgrounds have relatively soft spectra compared to all but the softest models considered.
For example, for a power law index of 1 when analyzing the full energy range, the upper limits are roughly 10 times lower than for an index of 2.2.

As can be seen from Table 3, the different dwarfs give roughly 
similar gamma-ray flux upper limits, as expected given the fairly uniform 
Fermi exposure across the sky.  However, the limits do vary from dwarf to 
dwarf due to, for example, the direction dependence of the diffuse background intensity and 
the proximity of bright gamma-ray sources.  In 
general, Ursa Minor gives the lowest flux limits while Sculptor, which is 
within a couple of degrees of a bright background point source, gives the 
highest flux limits.

\begin{deluxetable}{lccccc|cccccc}
\tabletypesize{\scriptsize}
\tablecaption{Flux Upper Limits at $95\%$ C.L.}
\tablewidth{0pt}
 \tablehead{
      & \multicolumn{5}{c}{ $E>100$ MeV}     & \multicolumn{5}{c}{ $E>1$ GeV} \\
 \colhead{Spectral index $\Gamma$}   & \colhead{1.0} & \colhead{1.8} & \colhead{2.0} & \colhead{2.2} & \colhead{2.4}   & \colhead{1.0} & \colhead{1.8} & \colhead{2.0} & \colhead{2.2} & \colhead{2.4} 

 }
\startdata
Ursa Major II   & 0.14 & 1.28 & 2.15 & 3.41 & 5.12 & 0.09 & 0.23 & 0.29 & 0.33 & 0.37 \\
Segue 2         & 0.10 & 0.71 & 1.28 & 2.33 & 4.21 & 0.06 & 0.12 & 0.15 & 0.17 & 0.19 \\
Willman 1       & 0.14 & 1.63 & 3.02 & 5.22 & 8.39 & 0.11 & 0.34 & 0.40 & 0.44 & 0.47 \\
Coma Berenices  & 0.08 & 0.43 & 0.69 & 1.11 & 1.74 & 0.05 & 0.07 & 0.08 & 0.09 & 0.09 \\
Bootes II      & 0.13 & 0.77 & 1.19 & 1.77 & 2.53 & 0.08 & 0.13 & 0.14 & 0.14 & 0.15 \\
Bootes I        & 0.12 & 1.02 & 1.71 & 2.70 & 3.96 & 0.09 & 0.23 & 0.28 & 0.31 & 0.33 \\
Ursa Minor      & 0.08 & 0.39 & 0.60 & 0.88 & 1.26 & 0.05 & 0.08 & 0.09 & 0.10 & 0.11 \\
Sculptor        & 0.22 & 2.43 & 3.88 & 5.71 & 7.76 & 0.12 & 0.34 & 0.39 & 0.44 & 0.48 \\
Draco           & 0.09 & 0.59 & 0.94 & 1.41 & 1.94 & 0.06 & 0.13 & 0.16 & 0.21 & 0.26 \\
Sextans         & 0.09 & 0.56 & 0.97 & 1.62 & 2.55 & 0.06 & 0.10 & 0.13 & 0.16 & 0.21 \\
Ursa Major I    & 0.09 & 0.48 & 0.77 & 1.23 & 1.90 & 0.06 & 0.09 & 0.10 & 0.12 & 0.14 \\
Hercules        & 0.33 & 1.51 & 2.22 & 3.23 & 4.63 & 0.24 & 0.30 & 0.30 & 0.28 & 0.27 \\
Fornax          & 0.12 & 0.94 & 1.72 & 3.05 & 5.04 & 0.09 & 0.14 & 0.16 & 0.17 & 0.18 \\
Leo IV          & 0.12 & 0.96 & 1.58 & 2.47 & 3.64 & 0.08 & 0.21 & 0.26 & 0.32 & 0.37 \\
\enddata
\tablecomments{Flux upper limits are given in units of $10^{-9} \mbox{cm}^{-2}\mbox{s}^{-1}$, above 100 MeV and 1GeV,  
 for the power-law model eq.(\ref{pwl}).}
\label{dwarf_DM_UL}
\end{deluxetable}

\section{Dark Matter constraints from dSph observations with the \emph{Fermi}-LAT detector}\label{models}

\subsection{Modeling of the dark matter density profiles}\label{profiles}

While power-law spectra can be justified on astrophysical grounds, 
a proper search for a Dark Matter signal should take account of the specific spectrum resulting
from WIMP annihilations.
At a given photon energy $E$, the $\gamma$-ray flux originating from
WIMP particle annihilations with a mass $m_{WIMP}$ can be
factorized into two contributions \citep{2008JCAP...07..013B}: the ``astrophysical factor''
$J(\psi)$ related to the density distribution in the emission region and
the ``particle physics factor'' $\Phi^{PP}$ which 
depends on the candidate particle characteristics :
\begin{equation}
\phi_{WIMP}(E,\psi)=J(\psi)\times \Phi^{PP}(E)\ ,
\label{def_phi}
\end{equation}
where $\psi$ is the angle between the direction of observation and the dSph center (as given in Table \ref{dwarfList}).
Following notations of \cite{2008JCAP...07..013B}, $J(\psi)$ and
$\Phi^{PP}$ are defined as
\begin{equation}
\Phi^{PP}(E) =\frac{1}{2}\frac{<\sigma v>}{4\pi\ m_{WIMP}^2} \sum_f \frac{dN_f}{dE} B_f,
\label{def_phiPP}
\end{equation}
and
\begin{equation}
J(\psi)=
  \int_{\rm{l.o.s}} dl(\psi) \rho^2(l(\psi)),
\label{def_Jpsi}
\end{equation}
where $<\sigma v>$ is the relative velocity times the annihilation cross-section of the two dark matter particles, 
averaged over their velocity distribution,
and the sum runs over all possible pair annihilation final states $f$, with 
$dN_f/dE$ and $B_f$ the corresponding photon spectrum and branching ratio, respectively. 
The integral in Eq.(\ref{def_Jpsi}) is computed
along the line of sight (l.o.s) in the direction $\psi$, and the integrand $\rho(l)$ is the assumed 
mass density of dark matter in the dSph.

For each galaxy, we model the dark matter distribution with a
Navarro-Frenk-White (NFW) \citep{Navarro:1996gj} density profile
within the tidal radius, as is reasonable for cold dark matter
sub-halos in Milky Way-type host halos \citep{2007ApJ...667..859D,
2008MNRAS.391.1685S}:
\begin{equation}
\rho(r) = \left\{ \begin{array}{ll}
	{{\rho_s r_s3} \over {r}(r_s+r)2} & \rm{for} \quad r < r_t \\
	0 & \rm{for} \quad r \geq r_t \\
\end{array} \right.,
\end{equation}
where $\rho_s$ is the characteristic density, $r_s$ is the scale
radius, and $r_t$ is the tidal radius. The line-of-sight integral in
Eq.(\ref{def_Jpsi}) may be computed once $\rho_s$, $r_s$ and $r_t$ are
known.
The tidal radius of the dwarf's dark matter halo in the gravitational
potential of the Milky Way is self-consistently computed from the Jacobi
limit~\citep{BinneyTremaine87} for each set of $\rho_s$ and $r_s$ values
assuming a mass profile for the Milky Way given
by~\citet{2009arXiv0907.0018C}. The sharp cut-off in the density
profile is rather extreme, but it is conservative
in the sense that it truncates the probability distribution of
expected $\gamma$-ray fluxes at the high end.

The halo parameters ($\rho_s$, $r_s$), and the resulting $J$ factor
from Eq.(\ref{def_Jpsi}), are estimated  following the methodology
outlined in \citet{Martinez09}.
The observed line-of-sight (l.o.s.) stellar velocities are
well-described by a Gaussian
distribution~\citep{Munoz05,Munoz06,Walker07,Walker09,Geha09}  and we
include the dispersion arising from both the motion of the stars and
the measurement errors as~\citet{Strigari:2006rd}:
\begin{equation}
\label{eq:fulllike}
\mathcal{L}(\mathscr{A}) \equiv P(\{v_i\}| \mathscr{A}) =  \prod_{i=1}^{n}
\frac{1}{\sqrt{2\pi(\sigma_{los, i}^2 + \sigma_{m, i}^2)}}
\exp \left[-\frac{1}{2}
\frac{(v_i -u)^2}{\sigma_{los, i}^2 + \sigma_{m, i}^2}
\right]\ ,
\end{equation}
where $\{v_i\}$ are the individual l.o.s. stellar velocity measurements
and $\sigma_{m,i}$ are the measurement errors on these velocities. The
mean l.o.s. velocity of the dwarf galaxy is denoted by $u$. The full
set of astrophysical parameters is
$\mathscr{A}={\rho_s,r_s,\Upsilon_\star,\beta,u}$, and we discuss the
two new parameters $\Upsilon_\star$ and $\beta$ below. The
theoretical l.o.s. dispersion, $\sigma_{los}$, is the projection of
the 3D velocity dispersion on the plane of the sky and this is
determined using the Jeans equation \cite[see][]{BinneyTremaine87}
once $\mathscr{A}$ is specified.  $\Upsilon_\star$ is the stellar mass
to light ratio and it sets the mass of the baryons in these dwarf
galaxies given the stellar luminosity. The velocity  dispersion
anisotropy is $\beta \equiv 1 - \sigma_t^2/\sigma_r^2$, where $\sigma_t$
and $\sigma_r$ are the tangential and radial velocity dispersion of
the stars (measured with respect to the center of the dwarf galaxy).
We assume that $\beta$ is constant for this analysis.  The probability
of the astrophysical parameters, $\mathscr{A}$ given a data set
$\{v_i\}$ is obtained via  Bayes' theorem:  $\mathcal{P}
(\mathscr{A}|\{v_i\}) \propto \mathcal{P}(\{v_i\}|\mathscr{A})
\mathcal{P}(\mathscr{A})$.  The prior probability,
$\mathcal{P}(\mathscr{A})$, for the halo parameters, $\{r_s, \rho_s\}$
is based on $\Lambda$CDM simulations \citep{2007ApJ...667..859D,
2008MNRAS.391.1685S} and described in detail in \citet{Martinez09}.
For $\Upsilon_\star$ we take the prior to be uniform between $0.5$ and
$5$, and for $\beta$ the prior is uniform between $-1$ and $1$. 

The
astrophysical factor $J$ after marginalization over all the parameters
in $\mathscr{A}$ for each dwarf galaxy within an angular region of
diameter 1$^\circ$ is given in Table \ref{DMdwarf}. The chosen 1$^\circ$ region for
the calculation of $J$ is a good match to the LAT PSF at energies of $1-2$ GeV where most
of the models under consideration are best constrained.  At lower energies, the PSF is significantly larger, but
beyond 1 $^\circ$ the dwarf dark matter density has a negligible impact on the overall J computation, and at higher energies, the statistics with the current data are rather limited.  Note that, due to
their uncertain nature as true dark matter dominated dSphs or large
uncertainties in their dark matter content, the Segue 2, Willman 1, and Bootes
II dSphs have not been considered in this analysis. In addition, new
stellar data on Segue 1 and Bootes II are being currently reduced and
will be used in a forthcoming publication. We also exclude Ursa Major
I, Hercules, and Leo IV, because their $J$ values are smaller than
those of the rest of the sample, yielding a final sample of 8 dSphs
used for the dark matter constraints.
\begin{deluxetable}{lccc}
\tabletypesize{\scriptsize}
\tablecaption{Properties of the dark matter halos of dwarf spheroidals used in this study.}
\tablewidth{0pt}
\tablehead{
\colhead{Name}               &  \colhead{[$\langle R \rangle$, $\langle P \rangle$]} &
\colhead{$\left[\langle R^2 \rangle - \langle R \rangle^2, \langle P^2 \rangle -
  \langle P \rangle^2, \langle R P \rangle - \langle R \rangle\langle P
  \rangle\right]$} & \colhead{$J^{NFW}$} \\ 
   \colhead{}                & \colhead{}  & \colhead{$R \equiv \log_{10}(r_s/kpc)$, $P \equiv
                   \log_{10}(\rho_s/M_\odot\ kpc^{-3})$} & \colhead{($10^{19} \frac{GeV^2}{cm^{5}}$)}
 }
 \startdata
Ursa Major II      & [$-0.78$, $8.54$] & [$0.0417$, $0.0986$, $-0.0554$] & $0.58_{-0.35}^{+0.91}$  \\
Coma Berenices      & [$-0.79$, $8.41$] & [$0.0603$, $0.132$, $-0.0820$] & $0.16_{-0.08}^{+0.22}$ \\
Bootes I            & [$-0.57$, $8.31$] & [$0.0684$, $0.165$, $-0.0931$] & $0.16_{-0.13}^{+0.35}$  \\
Usra Minor          & [$-0.19$, $7.99$] & [$0.0430$, $0.116$, $-0.0697$] & $0.64_{-0.18}^{+0.25}$ \\
Sculptor            & [$-0.021$, $7.57$] & [$0.0357$, $0.0798$, $-0.0528$] & $0.24_{-0.06}^{+0.06}$  \\
Draco               & [$0.32$, $7.41$] & [$0.0236$, $0.0364$, $-0.0286$] & $1.20_{-0.25}^{+0.31}$ \\
Sextans             & [$-0.43$, $7.93$] & [$0.0302$, $0.109$, $-0.0570$] & $0.06_{-0.02}^{+0.03}$	 \\
Fornax              & [$-0.24$, $7.82$] & [$0.0474$, $0.140$, $-0.0798$] & $0.06_{-0.03}^{+0.03}$	 \\
\enddata
\tablecomments{These parameters are obtained from measured stellar
  (line of sight) velocities. $\rho_s$ and $r_s$ are the density and
  scale radius for the dark matter halo distribution.  The first
  column, [$\log_{10}(\rho_s)$, $\log_{10}(r_s)$], is the average in the joint
  $\log_{10}(r_s)-\log_{10}(\rho_s)$ parameter space, whose posterior is well described by a Gaussian
distribution centered on the average value given.  The second
  column gives the diagonal and off diagonal components of the
  covariance matrix that may be used to approximate the joint
  probability 
  distribution of $\rho_s$ and $r_s$ as a Gaussian in $\log_{10}(r_s)$ and
  $\log_{10}(\rho_s)$. The last column provides $J^{NFW}$ (see Eq. \ref{def_Jpsi}), which is proportional to the pair
  annihilation flux coming from a cone of solid angle $2.4 ~ 10^{-4}$
  sr centered on the dwarf.  The errors on $J^{NFW}$ are obtained from
  the full MCMC probability distribution and bracket the range which
  contains $68\%$ of the total area under the probability
  distribution. } 
\label{DMdwarf}
\end{deluxetable}

In principle, annihilations in cold and dense substructure in the
dwarf galaxy halo can increase $J$. However, previous studies have
shown that this boost due to annihilations in substructure is unlikely
to be larger than a factor of few \citep[see e.g.][]{Martinez09}.  Similarly, a boost in the annihilation cross-section in dwarfs due to a Sommerfeld enhancement \citep[e.g.][]{2009PhRvD..79a5014A}, where the annihilation cross-section depends on the relative velocity of the particles, would increase the expected gamma-ray signal and improve our constraints.  In order to be conservative, we have not included either of these effects.

\subsection{Constraints on the annihilation cross-section}\label{DMconstraints}

Using Eq.(\ref{def_phi}), the results given in Table \ref{DMdwarf}, 
and the DMFit package \citep{2008JCAP...11..003J} as implemented in the 
{\tt  Science Tools},  95\% C.L. upper limits on photon fluxes (above 100 MeV) and 
on $<\sigma v>$ have been derived as a function of the WIMP mass,
 for each dSph and for specific annihilation channels.
Our choices for the pair-annihilation final states are driven by theoretical as well as phenomenological considerations: a prototypical annihilation final state is into a quark-antiquark pair. The resulting $\gamma$ rays stem dominantly from the decay of neutral pions produced in the quark and antiquark hadronization chains, and do not crucially depend upon the specific quark flavor or mass; in fact, a very similar $\gamma$ ray spectrum is produced by the (typically loop-suppressed) gluon-gluon final state. Here, for illustration we consider the specific case of a $b\bar b$ final state: this choice is motivated by the case of supersymmetric dark matter \citep[see][for a review]{Jungman:1995df}. In supersymmetry with R-parity conservation, the prototypical WIMP dark matter candidate is the lightest neutralino, the mass eigenstate resulting from the superposition of the fermionic partners of the hypercharge and SU(2) neutral gauge bosons and of the two neutral Higgs bosons. At moderate to large values of $\tan\beta$, if the lightest neutralino is bino-like (i.e. if the U(1) hypercharge gauge eigenstate is almost aligned with the lightest neutralino mass eigenstate), dark matter dominantly pair-annihilates into $b\bar b$. Another final state that is motivated by supersymmetric dark matter is into $\tau^+\tau^-$, that dominates in the case of a low-mass scalar superpartner of the $\tau$ lepton, as is the case e.g. in the so-called co-annihilation region of minimal supergravity (mSUGRA). An intermediate case with a mixed $b\bar b$ and $\tau^+\tau^-$ final state is also ubiquitous in supersymmetry, since those are the two heaviest down-type matter fermions in the Standard Model. The additional color factor and a larger value for the mass typically favor a relatively larger $b\bar b$ branching fraction.  While the choice of the final states we consider is motivated here by supersymmetry, the results we find apply to generic WIMP models and not only to neutralinos.

\begin{figure}
\begin{center}
\includegraphics*[width=8cm,height=8cm,angle=0]{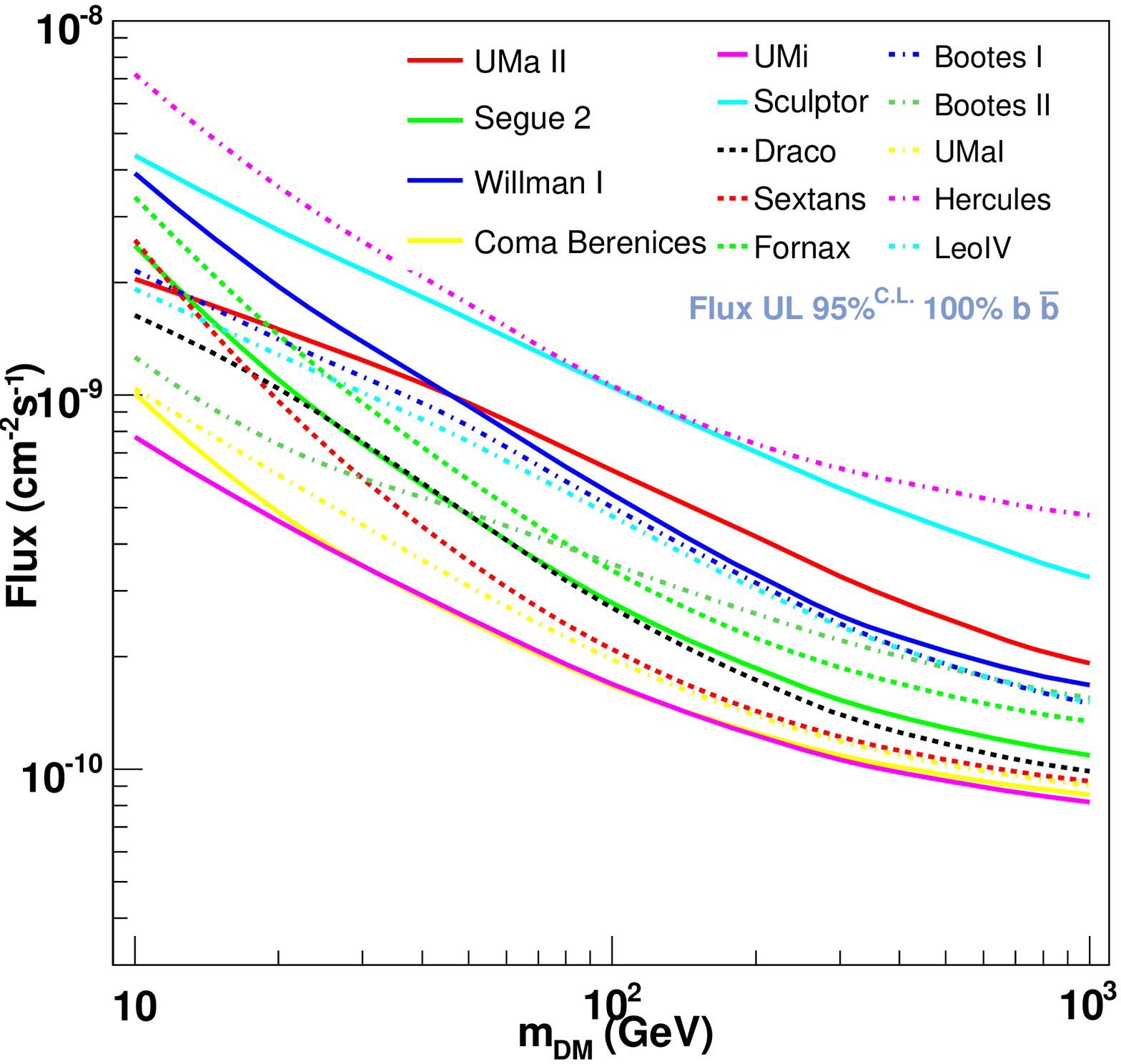}
\includegraphics*[width=8cm,height=8cm,angle=0]{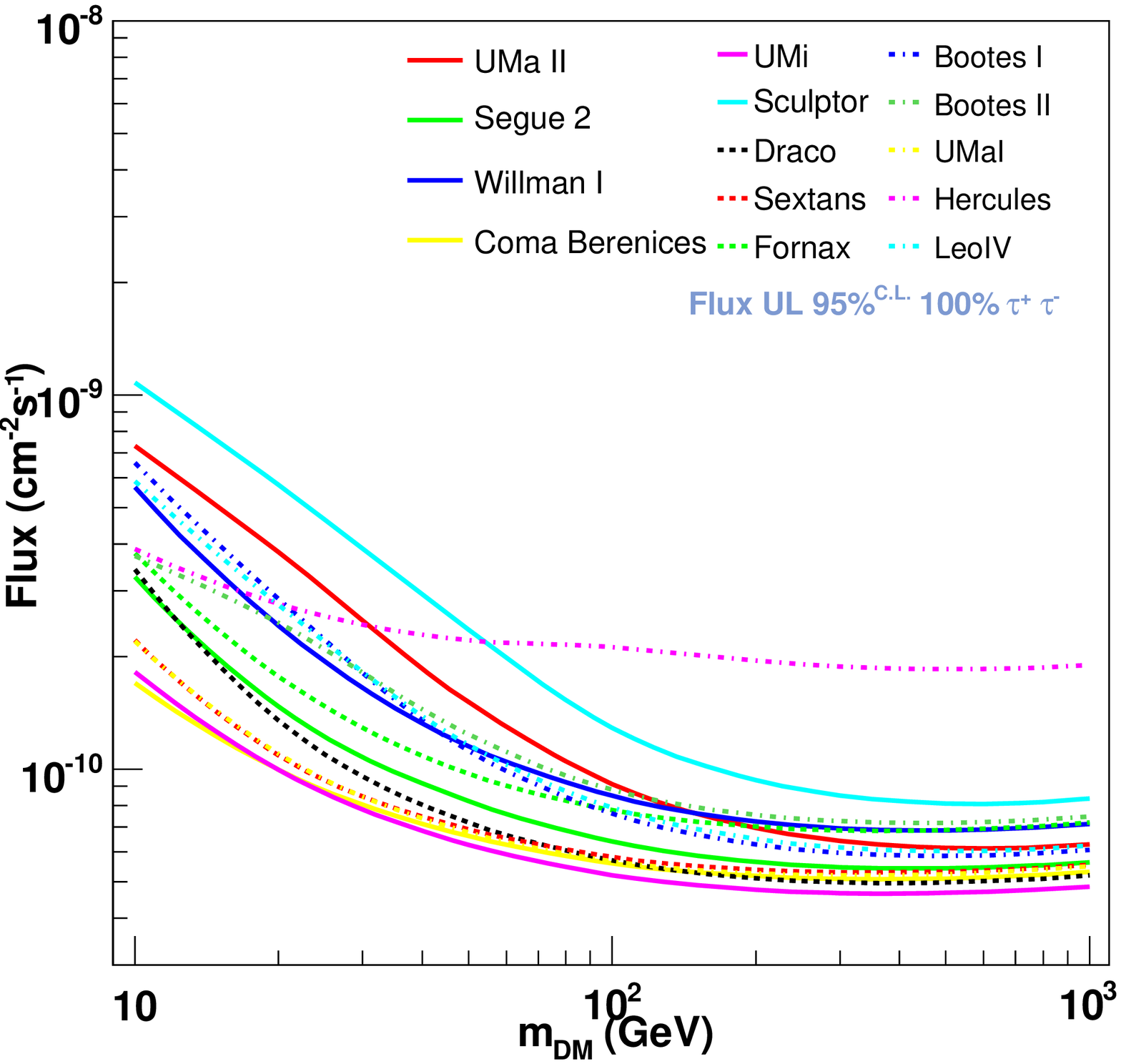}
\includegraphics*[width=8cm,height=8cm,angle=0]{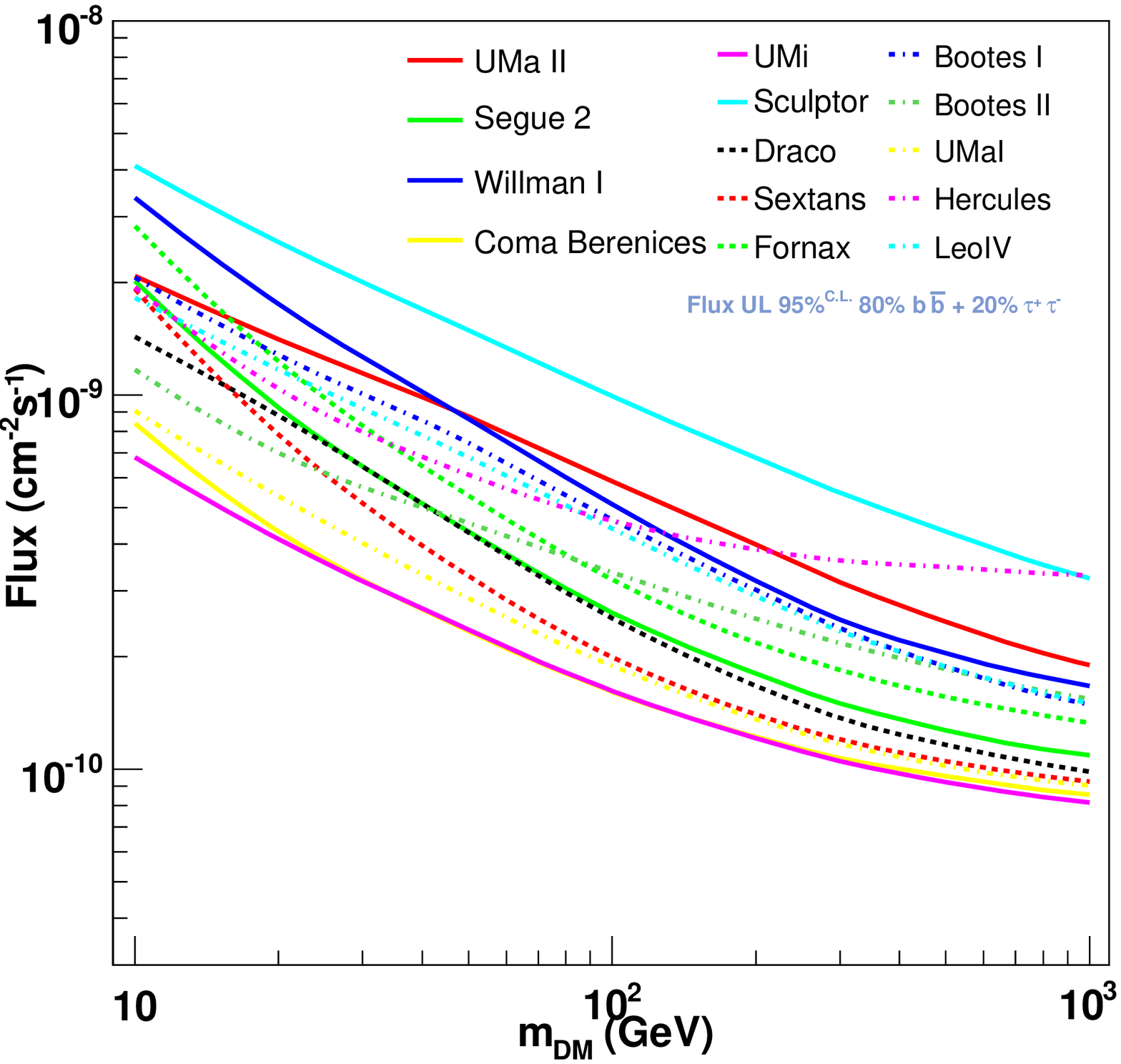}
\includegraphics*[width=8cm,height=8cm,angle=0]{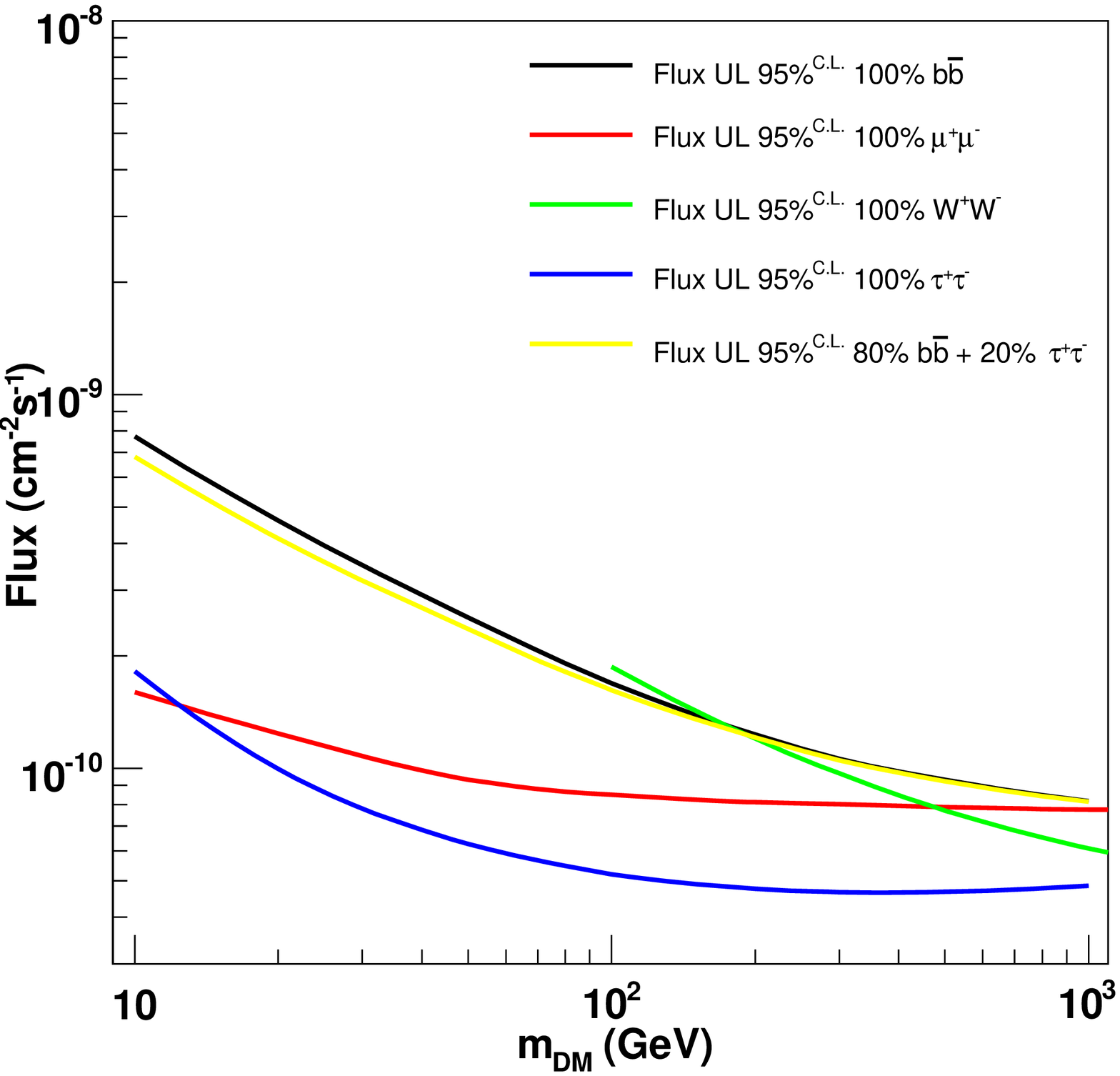}
\end{center}
\caption{\label{UL_Fluxes}
Derived upper limits on fluxes for all selected dwarfs and 
for various branching ratios: 
100$\%$ $b\bar{b}$ (upper left), 100$\%$ $\tau^+\tau^-$ (upper right) 
and mixed 80$\%\ b\bar{b}$ + 20$\%\tau^+\tau^-$ (lower left) final state.
Lower right plot gives an illustration of how the upper limits on the fluxes
can change depending on the selected final state (here for the Ursa Minor dSph).
}
\end{figure}

Fig.~\ref{UL_Fluxes} shows the
derived upper limits on the photon fluxes for all selected dwarfs and 
for various annihilation final states (respectively, 100$\%$ $b\bar{b}$ in the upper left panel, 100$\%$ $\tau^+\tau^-$ in the upper right panel, and a mixed 
80$\%\ b\bar{b}$ + 20$\%\tau^+\tau^-$ final state in the lower left panel).
The lower right plot of Fig.~\ref{UL_Fluxes} illustrates, for the case of Ursa Minor, 
how upper limits on the $\gamma$-ray fluxes change as a function of mass depending on the selected final state. Final states producing a hard $\gamma$-ray spectrum, such as $\mu^+\mu^-$ and $\tau^+\tau^-$ result in the best upper limits, since they predict abundant photons fluxes at larger energies, where the diffuse background is lower. With increasing mass, the spectrum from WIMP annihilation is rigidly shifted to higher energies, and the advantage of having a harder spectrum is less critical: for $m_{\rm DM}\sim 1$ TeV the upper limits we obtain for all the final states we consider vary within a factor 3, versus more than one order of magnitude at lower masses.

The results presented in fig.~\ref{UL_Fluxes} bracket realistic cases of theoretically motivated particle dark matter models, but since they only refer to the WIMP annihilation final state they do not depend on the particular assumed particle theory model.  We consider next a few motivated specific WIMP dark matter scenarios, and study how the corresponding parameter space is constrained.
We consider first the well-known case of mSUGRA (\cite{Chamseddine:1982jx, Barbieri:1982eh}; see also \cite{Nilles:1983ge} for an early review), where the supersymmetry breaking parameters are typically specified at a high-energy scale (typically taken to be the grand unification scale $M_{\rm GUT}\simeq2\times 10^{16}$ GeV) and assumed to be universal at that scale. Those parameters are the universal high-scale supersymmetry-breaking sfermion mass $m_0$, gaugino mass $M_{1/2}$ and trilinear scalar coupling $A_0$; an additional (low-scale) parameter is the ratio of the vacuum expectation values of the two Higgs doublets, $\tan\beta$, and the sign of the higgsino mass parameter $\mu$. Here, we adopt the following ranges for the various parameters (linear scan)~: 
$80 < m_{0} < 100000$,\
$80 < M_{1/2} < 4500$,\
$ A_{0}   = 0$,\
$1.5 < \tan\beta < 60$ and with sign($\mu$) being not constrained.

A less constrained alternative is to consider a ``phenomenological'' Minimal Supersymetric Standard Model (MSSM) setup \cite[see e.g.][for a review of the most general MSSM Lagrangian]{Chung:2003fi}, where all soft supersymmetry breaking parameters (i.e. the positive mass-dimension coefficients of lagrangian terms that explicitly break supersymmetry) are defined at the electro-weak (low-energy) scale, possibly with a few simplifying assumptions \citep[see e.g.][for an early attempt at a scan of the MSSM parameter space]{Profumo:2004at}. Here, we consider the reduced set of parameters considered in \cite{Gondolo:2004sc}, and perform a logarithmic scan over the following parameters~: 
$\vert {\mu} \vert < 10000$,\
$\vert M_{2} \vert < 10000$,\
$100 < m_{A} < 1000$,\
$1.001 < \tan\beta < 60$,\
$100 < m_{\tilde q} < 20000$, while the scan is linear in 
$-5 < A_{t}/m_{\tilde q} < 5$ and
$-5 < A_{b}/m_{\tilde q} < 5$.

In addition to the two scenarios considered above, we also entertain two additional specific WIMP dark matter models. The first one is the lightest Kaluza-Klein particle of Universal Extra Dimensions (UED) (see e.g. \cite{Cheng:2002ej, 2003NuPhB.650..391S}, for a review see \cite{Hooper:2007qk}), where, in the minimal setup, the dark matter candidate corresponds to the first Kaluza-Klein excitation of the $U(1)$ hypercharge gauge boson, also known as $B^{(1)}$. In this case, there is an almost fixed relationship between the dark matter mass and its pair annihilation cross section, and a thermal relic abundance in accord with the dark matter density is obtained for masses around 700 GeV \citep{2003NuPhB.650..391S}. 

The second model, which was recently considered in \cite{Kane:2009if} as a natural and well motivated scenario in connection with the anomalous positron fraction reported by PAMELA \citep{2009Natur.458..607A}, is that of  wino-like neutralino dark matter \citep{2000NuPhB.570..455M}. Wino-like neutralinos (which for brevity we will refer to as ``winos''), i.e. neutralino mass eigenstates dominated by the component corresponding to the supersymmetric fermionic partners of the SU(2) gauge bosons of the Standard Model, pair annihilate very efficiently into pairs of $W^+W^-$ (if their mass is larger than the $W$ mass) and the pair annihilation cross section is fixed by gauge invariance once the wino mass is given. Winos arise in various supersymmetry breaking scenarios and in several string motivated setups, where e.g. the anomaly mediation contribution to the gaugino masses dominates over other contributions, setting $M_2<M_1$. Typical such scenarios of anomaly mediated supersymmetry breaking (AMSB) were considered e.g. in \cite{1999NuPhB.557...79R, 1998JHEP...12..027G}. Although winos produce a thermal relic density matching the universal dark matter density for masses around 2 TeV, several non-thermal production mechanisms have been envisioned that could explain a wino dark matter scenario with lighter dark matter candidates than a TeV.

\begin{figure}
\begin{center}
\includegraphics*[width=8cm,height=8cm,angle=0]{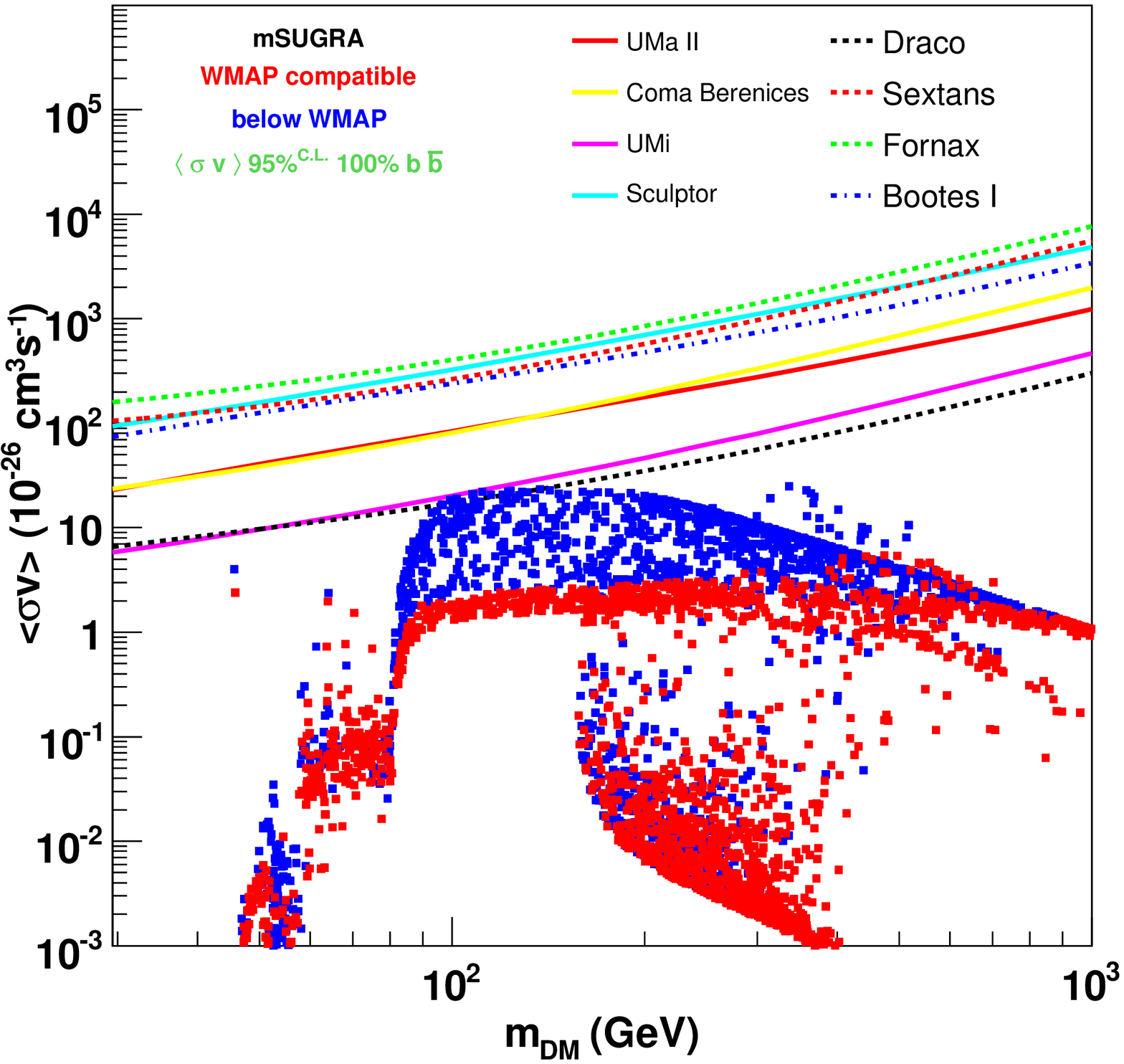}
\includegraphics*[width=8cm,height=8cm,angle=0]{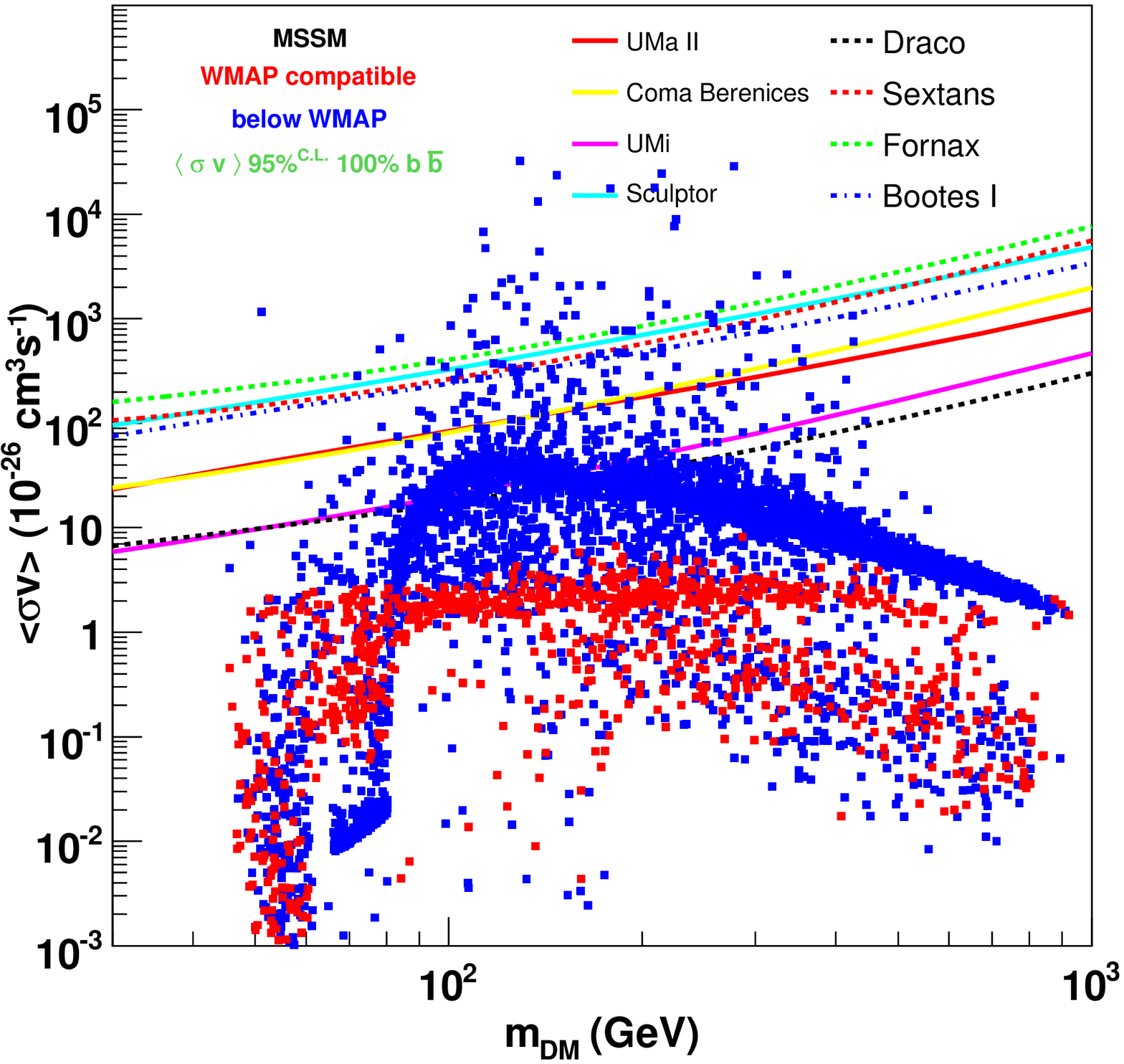}
\includegraphics*[width=8cm,height=8cm,angle=0]{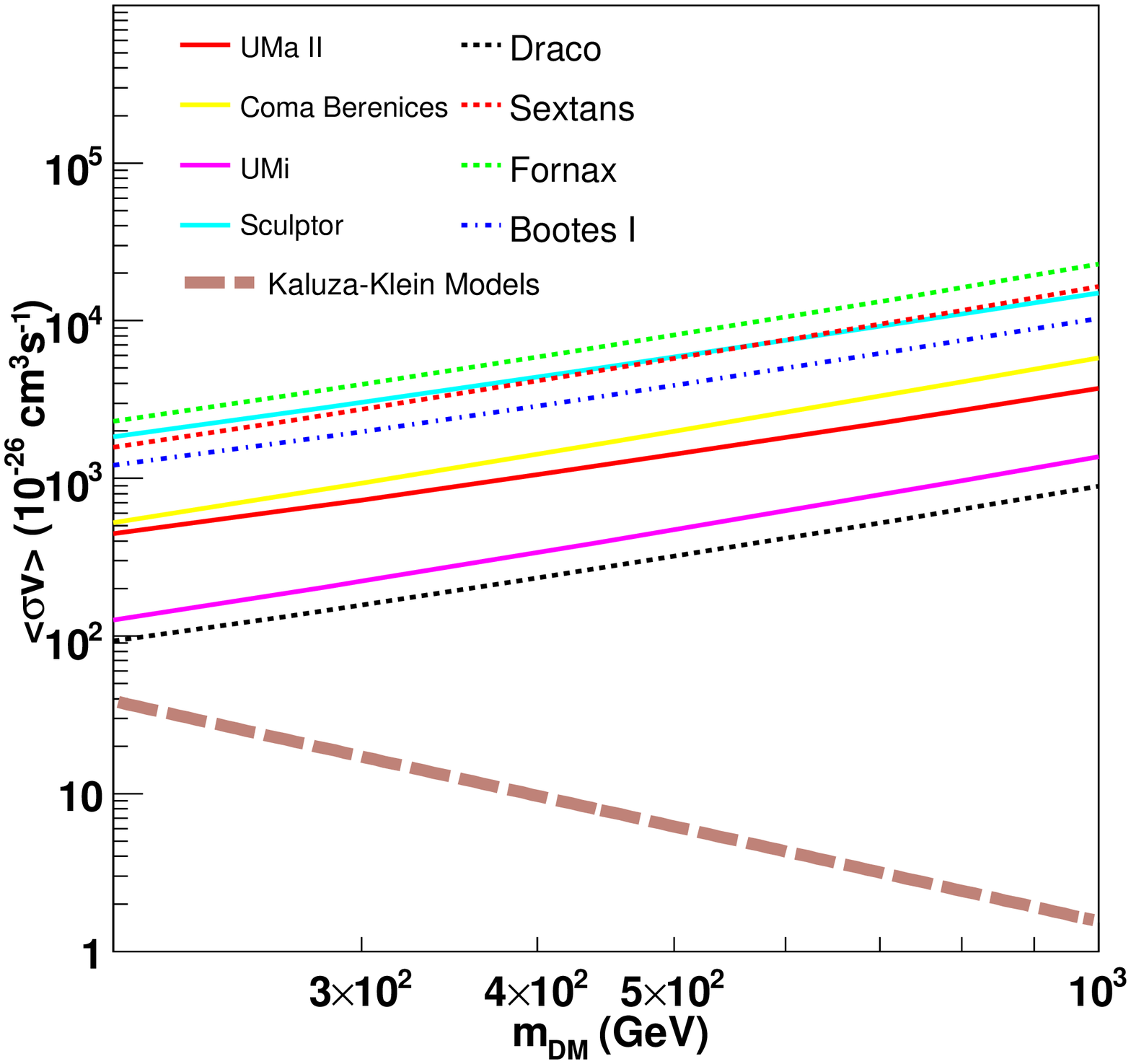}
\includegraphics*[width=8cm,height=8cm,angle=0]{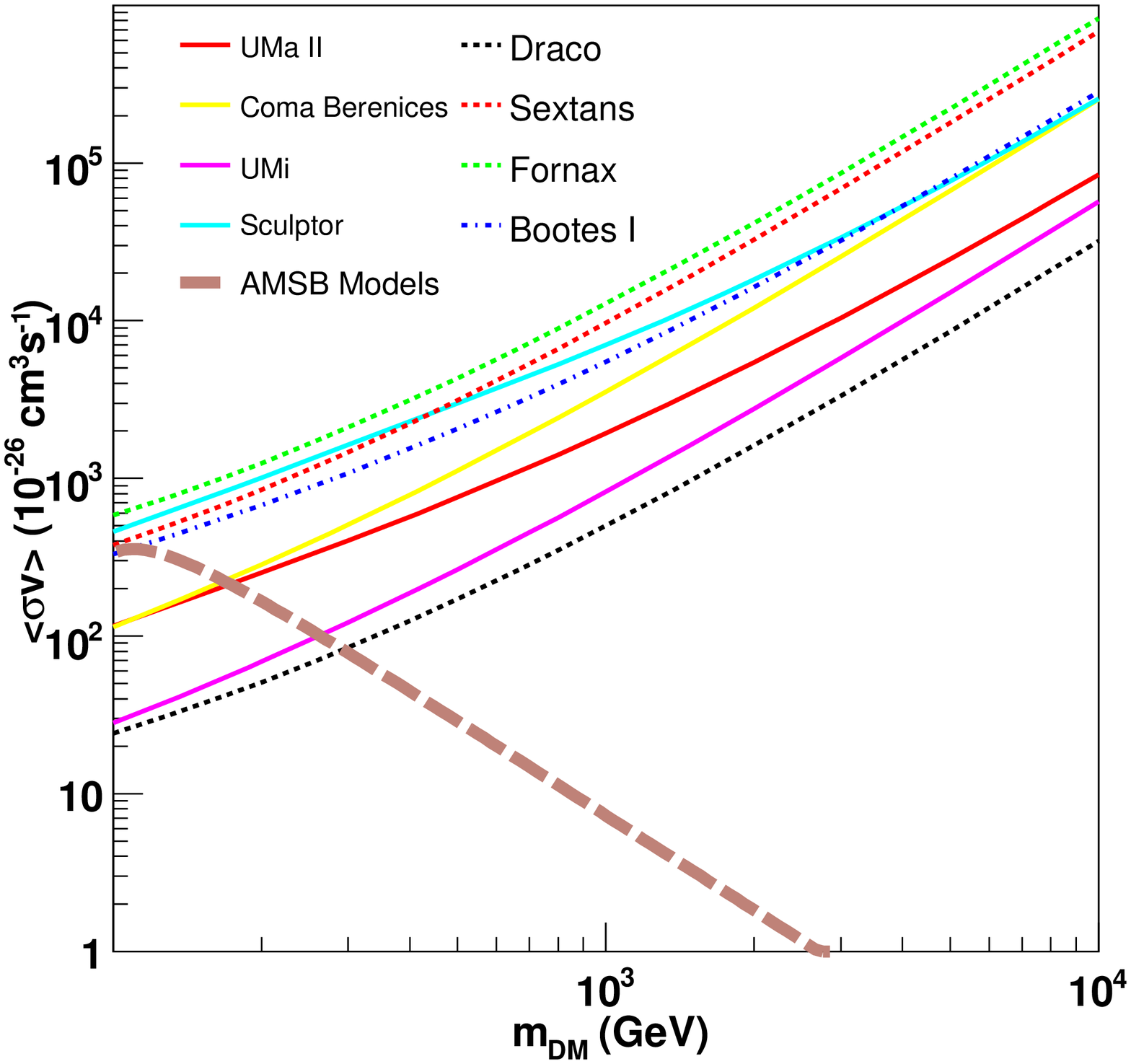}
\end{center}
\caption{\label{plots2}
mSUGRA (upper left), MSSM (upper right), 
Kaluza-Klein UED (lower left) and Anomaly mediated (lower right) 
models in the ($m_{{\rm wimp}}$,$<\sigma v >$) plane. 
All mSUGRA and MSSM plotted models are consistent with all accelerator constraints and
red points have a neutralino thermal relic abundance corresponding to the inferred cosmological dark matter density (blue points have a lower thermal relic density, and we assume that neutralinos still comprise all of the dark matter in virtue of additional non-thermal production processes).
The lines indicate the Fermi 
95\% upper limits obtained from 
likelihood analysis on the selected dwarfs given in Table \ref{DMdwarf}. 
}
\end{figure}

Fig.~\ref{plots2} compares the resulting LAT sensitivity in the ($m_{{\rm wimp}}$,$<\sigma v>$) plane 
with predictions from mSUGRA, MSSM, Kaluza-Klein dark matter in UED and wino-like dark matter in AMSB.
All mSUGRA and MSSM plotted models are consistent with accelerator constraints.
Red points are compatible with the 3$\sigma$ WMAP constraint on the universal matter density under thermal production while blue points would have a lower thermal relic density.  For the blue points, we assume that the production mechanism is non-thermal in order to produce the observed universal matter density, and we therefore do not rescale the neutralino density by the factor $(\Omega_{\rm thermal}/\Omega_{\rm DM})^2$, which would result assuming exclusive thermal production. This is very natural in the context of several string-theory motivated frameworks, where moduli generically decay into both Standard Model particles and their supersymmetric partners, which in turn eventually decay into the lightest neutralinos \cite{2000NuPhB.570..455M}. Topological objects such as Q-balls can also decay and produce neutralinos out of equilibrium, as envisioned e.g. by \cite{2002PhRvD..66h3501F} and \cite{2004PhRvD..69c5006F}. Another possible scenario is one where the expansion history of the Universe is more rapid than in a radiation dominated setup, for instance because of a dynamical ``Quintessence'' field in a kinetic-dominated phase \citep{2003PhLB..571..121S,2003JCAP...11..006P}.

Fig.~\ref{plots2} clearly shows that, after less than a year of \emph{Fermi} data survey, the upper limits on the $\gamma$-ray flux from dSphs are already starting to be competitive for MSSM models, provided that these models correspond to  low thermal relic density. Draco and Ursa Minor dSphs set the best limits so far. Pending more data, they may also start to constrain mSUGRA models with low thermal relic density as well.
Furthermore, these flux upper limits 
already disfavor AMSB
 models with masses $<$300 GeV. Interestingly, our results strongly constrain the models considered in \cite{Kane:2009if}, invoking a 200 GeV mass wino.

\subsection{Comparison to dark matter models proposed to fit the PAMELA and Fermi $\mathbf{e^+e^-}$ data}
\label{mumu}

\begin{figure}
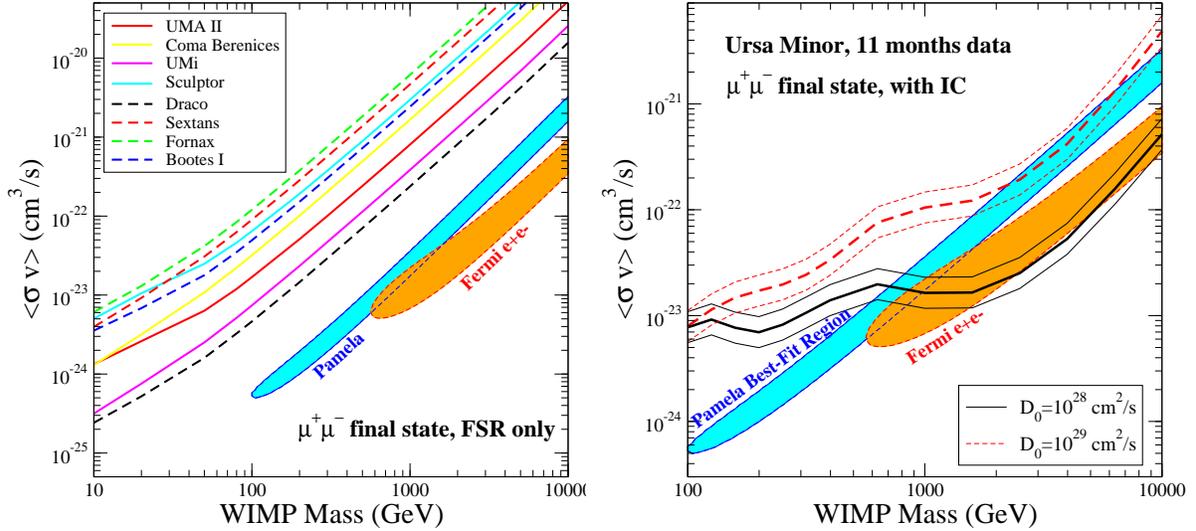

\begin{center}
\includegraphics[width=7.8cm,angle=0]{fsronly_newer.eps}
\includegraphics[width=7.8cm,angle=0]{ursaminor_new.eps}
\caption{\label{plots3}Constraints on the annihilation cross-section 
for a $\mu^+\mu^-$ final state based on the 95\% confidence limits on the $\gamma$-ray flux compared to dark matter annihilation models 
which fit well either the PAMELA \citep{2009Natur.458..607A} or Fermi $e^++e^-$ measurements \citep{2009PhRvL.102r1101A}.  The left panel shows the constraints considering $\gamma$-ray emission from final state radiation only.  The 
right panel
shows the constraints for the Ursa Minor dwarf including both $\gamma$-ray emission from IC 
scattering and final state radiation.  Here we consider two different diffusion coefficients, 
and show the effect of the uncertainties in the Ursa Minor density profile.}
\end{center}
\end{figure}

The recent detection by the PAMELA experiment of a positron fraction that increases with 
energy above 10 GeV \citep{2009Natur.458..607A} and the possibility that dark matter annihilation in the 
Galaxy could produce this ``positron excess" (among more mundane explanations such as pulsars) 
have spurred great interest in the particle physics community. The pair annihilation of galactic WIMP dark matter can, in principle, produce an anomalous excess in the positron fraction at energies between a few GeV and $\sim100$ GeV. The spectrum of high-energy $e^++e^-$, although compatible with a purely canonical cosmic-ray origin \citep{2009PhRvL.102r1101A,Grasso:2009ma}, can also accommodate an additional component due to galactic dark matter annihilation.

A dark matter annihilation interpretation of the positron excess implies preferentially a leptonic final state, to avoid the over-production of antiprotons. This is very hard to achieve in the minimal supersymmetric extension of the Standard Model \citep[see however ][for the case of AMSB, which we will not consider further here]{2009arXiv0906.4765K}. In addition, the spectral shape of high-energy $e^+e^-$ points towards rather large masses, and the level of the needed local positron flux indicates either a very large pair-annihilation rate, or a strong enhancement in the local dark matter density. Using a canonical primary electron injection spectrum, the analysis of \cite{2009PhRvL.103c1103B} further indicates that a preferred annihilation final state is $\mu^+\mu^-$, or the somewhat softer (in the produced $e^+e^-$ spectrum) but essentially very similar four body $\mu^+\mu^-\mu^+\mu^-$ final state. Theoretical arguments that could explain this peculiar annihilation final state, possibly involving mechanisms to enhance the low-velocity annihilation rate, have been proposed \citep{2009PhRvD..79a5014A,2009PhRvD..79g5008N}. With standard assumptions on the dark matter density profile, and assuming a $\mu^+\mu^-$ final state, the regions in the pair-annihilation cross section versus mass plane preferred by the \emph{Fermi}-LAT $e^+e^-$ data are shown in orange in Fig.~\ref{plots3}, while those favored by the PAMELA positron fraction data are highlighted in light blue (for details on the computation of these regions see \cite{2009JCAP...07..020P}).

In a pair-annihilation event producing a $\mu^+\mu^-$ pair, $\gamma$ rays result from both the internal bremsstrahlung off of the muons (final state radiation), with the well-known hard power-law spectrum ${\rm d}n_\gamma/{\rm d}E_\gamma\sim E^{-1}_\gamma$, and from the inverse Compton (IC) up-scattering of cosmic microwave background light by the $e^+e^-$ resulting from muon decay.  The dark matter interpretation of the ``cosmic-ray lepton anomalies'' implies significant $\gamma$-ray emission from a variety of sources; predictions and constraints on these models have been discussed extensively in the recent literature \citep[see e.g.][for a discussion of the constraints from the expected IC emission from annihilation at all redshifts and in all halos]{2009JCAP...07..020P}.

The left panel of Fig.~\ref{plots3} illustrates the constraints we derive from 11 months of Fermi data on local dSph on  a generic WIMP dark matter pair-annihilating into a $\mu^+\mu^-$ final state (we do not specify here any particular particle physics scenario, although as stated above several examples have been considered in the literature), considering the final state radiation emission only. Here the spectral modeling was done using the DMFit package as in \S\ref{DMconstraints}.  The hierarchy among the constraints derived from the various dSph is very similar to what we find for the softer pair-annihilation final states considered in Fig.~\ref{plots2}. Neglecting any low-velocity enhancement of the annihilation rate \citep{2009PhRvD..79a5014A}, which would boost the $\gamma$-ray signal from dSph and hence the constraints we show, the final state radiation alone does not yet exclude portions of the parameter space favored by the dark matter annihilation interpretation of the cosmic ray lepton data.

The calculation of the $\gamma$-ray yield from IC is complicated, in systems as small as dSphs, by the fact that $e^+e^-$ are not confined (i.e. their diffusion lengths are typically larger than the physical size of the system). In fact, the typical energy-loss length-scale for TeV $e^+e^-$ loosing energy dominantly via IC off of photons in the cosmic microwave background is of the order of hundreds of kpc, much larger than the size of dSph. Assumptions on cosmic ray diffusion are therefore critical, as discussed e.g. in \cite{2007PhRvD..75b3513C} and in \cite{2008ApJ...686.1045J}. In the absence of any direct cosmic-ray data for external galaxies such as dSph (the only piece of information being that dSph are gas-poor environments with typically low magnetic fields), we consider the usual diffusion-loss equation and solve it in a spherically symmetric diffusive region with free-escape boundary conditions. We employ a diffusion coefficient at the level of what is usually inferred for cosmic rays in our own Milky Way \citep[e.g.][]{Strong:2007nh} (for a thorough discussion of the diffusion model we adopt here, we refer the reader to \cite{2007PhRvD..75b3513C} and  \cite{2008ApJ...686.1045J}). Specifically, we consider two values for the diffusion coefficient, $D_0=10^{28}\ {\rm cm}^2/{\rm s}$ and $D_0=10^{29}\ {\rm cm}^2/{\rm s}$ bracketing the values typically inferred for the Galaxy: the larger the diffusion coefficient, the larger the cosmic ray mean free path, and the larger the leakage of cosmic ray $e^+e^-$ out of the dwarf, leading to a suppression of the IC signal.  We also assume a power-law dependence of the diffusion coefficient on energy given by $D(E)=D_0\ \left(\frac{E}{\rm 1\ GeV}\right)^{1/3}$. In the energy loss term, IC emission off of Cosmic Microwave Background photons by far dominate over both synchrotron and starlight IC losses.

We show our results, {\em including both final state radiation (FSR) and IC emission off of CMB photons}, in Fig.~\ref{plots3}, right panel, for the case of the Ursa Minor dSph. Here the spectrum was modeled self-consistently with custom spectra
which include the expected $\gamma$-ray emission from both IC and FSR for a given assumed particle mass and diffusion coefficient for a grid of particle mass values ranging from 100 MeV to 10 TeV.  The Ursa Minor dSph was chosen as an example of one of the best cases for this particular study.  In the smaller ultra-faint dwarfs, diffusion is expected to have a much larger effect due to the much smaller diffusive region (modeled based on the stellar extent), and IC emission for the diffusion coefficients assumed here will not add significant flux above what is expected from FSR alone.  

The lower and upper lines in Fig.~\ref{plots3} indicate the range of uncertainty in the determination of the dark matter density profile of the Ursa Minor dSph. With the smaller diffusion coefficient choice, and for large enough masses (producing higher energy $e^+e^-$ and subsequent IC photons), the IC emission dominates, and it exceeds our $\gamma$ ray upper limits for models that fit the PAMELA data and have masses larger than 1 TeV. Excessive emission is predicted also in the more conservative case with $D_0=10^{29}\ {\rm cm}^2/{\rm s}$ for some models with masses in the 2-5 TeV range.

\section{Conclusions and final remarks}
\label{conclusion}

We have reported the observations of $\gamma$-ray
emission from 14 known dwarf spheroidal galaxies by \emph{Fermi}-LAT.
No excesses have been observed in LAT data and upper limits have been
derived on the $\gamma$-ray flux from dSphs.

Using the dark matter halo modeling for the 8 best candidate dwarf 
spheroidal galaxies derived from the latest stellar data (tab.~\ref{DMdwarf}), 
we have shown that if dark matter is assumed to consist entirely of
neutralinos, the upper limits obtained from one year of LAT data begin
to constrain mSUGRA and MSSM models with low thermal relic densities and AMSB models with wino-like neutralinos with masses below 300 GeV (fig.~\ref{plots2}). It is worth noting that four dSphs have also been observed by Cherenkov telescopes : Sagittarius by H.E.S.S. \citep{2008APh....29...55A}, Draco and Ursa Minor by Whipple \citep{2008ApJ...678..594W} and Veritas \citep[][which also includes Willman 1]{2009arXiv0910.4563W}, and finally Draco \citep{2008ApJ...679..428A} and Willman 1 \citep{2009ApJ...697.1299A} by MAGIC. The observation time varies between these studies, but in general the limits on the annihilation cross-section reported vary between a few times $10^{-23}$ to a few times $10^{-22}$ cm$^{-3}$ s$^{-1}$ for a 1 TeV mass neutralino and an assumed NFW dwarf density profile. IACT observations are most sensitive to typically higher mass dark matter particles (greater than $\sim$ 200 GeV) compared to the LAT, making them complimentary to Fermi searches.

The \emph{Fermi} limits also constrain WIMP models proposed to explain the \emph{Fermi} and PAMELA $e^+e^-$ data, 
particularly for high particle masses ($>$ 1 TeV, fig.~\ref{plots3}).  For these models,
strong constraints come from the inclusion of the expected IC $\gamma$-ray 
emission, though the flux of this component depends on the assumed diffusion 
model of $e^+e^-$ in dSphs.  

It is worth emphasizing that the results presented in this paper have all been obtained 
for a standard NFW halo shape and without assuming any boost
factor effect due to substructures in the dwarfs or a Sommerfeld enhancement to the annihilation cross-section.

\acknowledgments

We would like to thank the referee for their valuable comments and improvements to the paper.  Extended discussions with M. Geha, J. Simon, L. Strigari, and J. Siegal-Gaskins are gratefully
ackowledged.  The \textit{Fermi} LAT Collaboration acknowledges generous ongoing support
from a number of agencies and institutes that have supported both the
development and the operation of the LAT as well as scientific data analysis.
These include the National Aeronautics and Space Administration and the
Department of Energy in the United States, the Commissariat \`a l'Energie Atomique
and the Centre National de la Recherche Scientifique / Institut National de Physique
Nucl\'eaire et de Physique des Particules in France, the Agenzia Spaziale Italiana
and the Istituto Nazionale di Fisica Nucleare in Italy, the Ministry of Education,
Culture, Sports, Science and Technology (MEXT), High Energy Accelerator Research
Organization (KEK) and Japan Aerospace Exploration Agency (JAXA) in Japan, and
the K.~A.~Wallenberg Foundation, the Swedish Research Council and the
Swedish National Space Board in Sweden.

Additional support for science analysis during the operations phase is gratefully
acknowledged from the Istituto Nazionale di Astrofisica in Italy and the Centre National d'\'Etudes Spatiales in France.

\bibliography{LAT_Biblio,GC_Biblio,DM_Biblio,DWARF_Biblio,Martinez,BeyonSM,Stat}


\end{document}